\documentstyle{article}
\newfont{\bg}{cmr10 scaled\magstep4}

\newcommand{\bigzerou}{%
\smash{\lower1.7ex\hbox{\bg 0}}}
\newtheorem{theorem}{Theorem}
\newtheorem{prop}{Proposition}
\newtheorem{defi}{Definition}
\newtheorem{cor}{Corollary}
\newtheorem{conj}{Conjecture}
\newtheorem{lem}{Lemma}

\begin{document}
\title{
\begin{flushright}
  \begin{minipage}[b]{5em}
    \normalsize
      UT-HEP-760-96\\
  \end{minipage}
\end{flushright}
{\bf On the Structure of the Small Quantum Cohomology Rings \\
of Projective Hypersurfaces}}
\author{Alberto Collino$\;^{1)}$, Masao Jinzenji$\;^{2)}$\\
\\
\it 1) Dipartimento di Matematica, Universita' di Torino\\
\it Via Carlo Alberto 10, 10123 Torino, Italy\\
\it 2)Department of Physics, University of Tokyo\\
\it  Bunkyo-ku, Tokyo 113, Japan}
\maketitle
\begin{abstract}
We give an explicit
procedure which computes
for degree $d \leq 3$ the correlation
functions of topological sigma model (A-model)
on a projective Fano hypersurface $X$ as
homogeneous polynomials of degree $d$ in the  correlation
functions of degree 1 (number of lines).
We  extend this formalism to the case
of Calabi-Yau hypersurfaces and
explain how the polynomial property
is preserved.
Our key tool is the construction of 
universal recursive formulas
which express the structural constants of
the  quantum cohomology ring of $X$
as weighted homogeneous polynomial functions in the constants of the
Fano hypersurface with the same degree and dimension one more.
We propose some conjectures about the existence
and the form of the recursive formulas for
the structural constants of rational curves
of arbitrary degree. Our recursive formulas should yield the 
coefficients of  the hypergeometric series
used in the mirror calculation. Assuming the validity of the
conjectures  we find the 
recursive laws
for rational curves of degree $4$ and $5$.
\footnote{ e-mail
address: collino@dm.unito.it,  jin@danjuro.phys.s.u-tokyo.ac.jp}
\end{abstract}

\section{Introduction}
In \cite{jin}, we studied the K\"ahler sub-ring
$H^{*}_{q,e}(M_{N}^{k})$ in the quantum
cohomology ring of a 
hypersuface $M_{N}^{k}$ of degree $k$  in $CP^{N-1}$,
we used numerical computation based on the
torus action method.  We worked under the condition 
that $c_{1}(M_{N}^{k})$ is not negative,
i.e. under the ipothesis $ N-k\geq 0 $. The following
statements summarize the content of that paper:

1.For $N\leq 9$ with
$N-k  \geq 2$, we 
computed  
that the main relation 
satisfied by the generator ${\cal O}_{e}$ of
$H^{*}_{q,e}(M_{N}^{k})$
 has the simple form
\begin{equation} 
({\cal O}_{e})^{N-1}-k^{k}({\cal O}_{e})^{k-1}\cdot q=0
\label{main}
\end{equation}

2. Under the same restriction as for 1
the structural constants  of $H^{*}_{q,e}(M_{N}^{k})$ can be
expressed as polynomial functions of a finite set of
integers. These integers are the Schubert numbers of lines,
they do depend from the degree 
of the hypersurface but not from its dimension.

3. An explanation for (\ref{main}) 
was found by looking at a
toric compactification of the moduli space of maps from
$CP^{1}$ to $CP^{N-1}$. It was said that 
the boundary portion of the moduli
space should turn out
to be irrelevant for the calculation,
under the condition $ N-k \geq 2$

The justification for 1. and 2. was
based on numerical computations  
and the explanation of 3. was heuristic.

Givental \cite{givental} gave a mathematically
rigorous proof  of (\ref{main}). He constructed the exact solution
of the Gauss-Manin system ( the deformation parameter is restricted to
the
K\"ahler deformation) associated to A-model on $M_{N}^{k}$
by using torus action method and he showed that it satisfies
the linear ODE of hypergeometric type if $N-k\geq 2$.
This ODE reduces to the relation of $H^{*}_{q,e}(M_{N}^{k})$ under
certain limit (in his notation $\hbar\rightarrow 0$) and we have
(\ref{main}). He also treated the cases when $N-k$ is $1$ or $0$ and
he showed that the solution above  satisfies the linear ODE of
hypergeometric type if

a) some multiplicative factor are added. (when $N-k=1$)

b) some multiplicative factor are added and at the same time,
coordinate transformation
(by mirror map) is performed. (when $N-k=0$)

With b) Givental proved the mirror symmetry conjecture,
namely that topological sigma models on Calabi-Yau manifolds realized
as the complete intersections of $CP^{N-1}$ can be solved by the
analysis of hypergeometric series. His proof of the symmetry
seems to rely on the flat metric condition or the fact that the
three point 
functions including identity operator do not receive quantum
correction. Then the argument goes very smoothly
but we can hardly see what is happening microspically in compensation
for the smoothness.
In this paper   we try
to explain (\ref{main}) by descending
induction of $N$. Our program is to  construct
recursive formulas that express the structural constants of
$H^{*}_{q,e}(M_{N}^{k})$ as weighted homogeneous
polynomials in the structural constants of
$H^{*}_{q,e}(M_{N+1}^{k})$.
Our method is based on a geometric process,
which we call the  specialization procedure, unfortunately it works
only up to the case of cubic curves.
We believe that it should be possible
to find and construct universal recursive
laws also for curves of higher degree.
We state some ansatz
on the expected structure of such formulas.
If the index $N-k \geq 2$,
the recursive formulas should stay the same,
independently of  $N$ and $k$,
and they must imply that 
the main relation if
$H^{*}_{q,e}(M_{N}^{k})$ is of the 
type given above  (\ref{main}). 
When the hypersurface is of Fano index
is $1$, then the resursion law for the 
Schubert numbers of lines changes, while
the formulas for curves of higher degree do not.  
Coming to the Calabi-Yau situation,
$N-k= 0$, the recursion relations 
is modified for all degrees. The main relation
of (\ref{main}) must be changed entirely.
We first computed the recursion relation for lines in case of
$H^{*}_{q,e}(M_{N}^{N})$ and evaluated the degree 1 part of the relation
\cite{jin}.The result had a structure strongly similar to the result
from
mirror symmetry \cite{mnmj} and we speculated that the above correction
and the correction terms argued in \cite{mnmj} are closely related.
On the other hand the universal recursion laws
valid for the case $N-k
\geq 2$  can be formally iterated by descent
of dimension (while keeping the degree of
the hypersurface fixed) up to
the case of a Calabi-Yau. What we conjecture here, and verify
in part, is that in this manner 
one recovers the coefficients
of the hypergeometric series which appear
in the mirror calculation, but we obtain
them without use of the mirror conjecture.
At this point the construction of the correction
terms for the quantum ring
can be  done by the procedure that arises from the flat metric
condition.

        We prove the main relation (\ref{main}) by 
means of the recursive formulas,
modulo ($q^4$). It is clear from the topological 
selection rule that when $N$
is large enough with respect to the degree $k$ then  the only
non null  quantum corrections left come from  lines.
We construct explicit recursion relations for $d \leq 3$ and
prove (\ref{main}) and 2. within this range
by decreasing induction on $N$. 
We think that the universal recursive procedure should
provide interesting
information also for the case when $M_{N}^{k}$ is a 
hypersurface of general type, i.e. $ N<k$.

This paper is organized as follows.

In section 2,  we recall first the main properties  of the structure of
the quantum K\"ahler algebra
$H^{*}_{q,e}(M_{N}^{k})$ and then we study  the quantum product 
with primitive cohomology classes for the Fano case, $ k<N$.

In section 3, we introduce the specialization calculation and derive
the recursion relations for rational  curves
of degree at most $3$ under the assumption that
the hypersurface is a Fano manifold.

In section 4, we extend the specialization procedure to 
Calabi-Yau hypersurfaces and determine
how the recursion relation should be modified (up to degree 2).
Using these results, we evaluate the
main relation of $H^{*}_{q,e}(M_{N}^{N})$ and compare it with
the result from mirror symmetry.
We will also show that our findings can be organized 
in a compact form by means of
the hypergeometric series used in mirror calculation.
This is our reason for a conjecture which says how to  
modify the 
recursion relation 
in the case of hypersurfaces of Calabi-Yau type.

In section 5, we present a set of conjectures, which
should provide a guiding rule in the explicit construction
of the recursive formulas for rational curves of
higher degree. Assuming the conjectures
we explicitly construct the recursive formulas for degree 4 and 5.
 
\section{Quantum cohomology of Fano hypersurfaces}
\subsection{ The quantum K\"ahler Sub-Ring $H^{*}_{q,e}(M_{N}^{k})$}
Let  $M_{N}^{k}$ be the  hypersurface of degree $k$ in $CP^{N-1}$.
By the Lefschetz theorem the cohomology ring $H^{*}(M_{N}^{k})$ splits
into two parts. One of them is the K\"ahler sub-ring generated by
the  K\"ahler form
$e$ induced from the hyperplane section
$H$ of $CP^{N-1}$, and the other is the primitive part, which is a
subspace of the middle dimension cohomology
$H^{N-2}(M_{N}^{k})$.
We first consider the quantum
K\"ahler sub-ring $H^{*}_{q,e}(M_{N}^{k})$, it is
generated additively by
${\cal O}_{e^{\alpha}}\quad (\alpha= 0,1,2,
\cdots N-2)$, where ${\cal O}_{e^{\alpha}}$ represents the
BRST- closed operator induced from $e^{\alpha}\in H^{*}(M_{N}^{k})$.
The multiplication rules of $H^{*}_{q,e}(M_{N}^{k})$ are
determined by means of the flat metric and the
three point correlation functions
(or Gromov-Witten invariants):
\begin{eqnarray}
\eta^{N,k}_{\alpha\beta}&:=&\langle{\cal O}_{e^0}{\cal O}_{e^{\alpha}}
{\cal O}_{e^{\beta}}\rangle_{M_{N}^{k}}=
\int_{M_{N}^{k}}e^{\alpha}\wedge e^{\beta}
=k\cdot\delta_{\alpha+\beta,N-2}\\
\eta^{\alpha\beta}_{N,k}\eta_{\beta\gamma}^{N,k}
&=&\delta^{\alpha}_{\gamma},\quad
\eta^{\alpha\beta}_{N,k} =\frac{1}{k}\cdot\delta_{\alpha+\beta,N-2}\\
C_{\alpha,\beta,\gamma}^{N,k}&=&\langle{\cal O}_{e^{\alpha}}
{\cal O}_{e^{\beta}}{\cal O}_{e^{\gamma}}\rangle_{M_{N}^{k}}=
\sum_{d=0}^{\infty}q^{d}\int_{{\cal M}_{0,d,3}^{M_{N}^{k}}}
\phi_{1}^{*}(e^{\alpha})\wedge \phi_{2}^{*}(e^{\beta})\wedge
\phi_{3}^{*}(e^{\gamma})\\
\phi_{i}:{\cal M}_{d,0,3}^{M_{N}^{k}}&\mapsto &M_{N}^{k},\quad
(\phi_{i}(\{z_{1},z_{2},z_{3}, f\}/\sim)= f(z_{i}))\nonumber\\
q&:=& e^{t}\nonumber.
\end{eqnarray}
The rules of quantum multiplication  are
\begin{equation}
{\cal O}_{e^{\alpha}}\cdot{\cal O}_{e^{\beta}}=
C_{\alpha,\beta,\gamma}^{N,k}\eta^{\gamma\delta}{\cal O}_{e^{\delta}}
=\frac{1}{k}C_{\alpha,\beta,\gamma}^{N,k}{\cal O}_{e^{N-2-\gamma}}:=
\sum_{d=0}^{\infty}q^{d}\frac{1}{k}C_{\alpha,\beta,\gamma}^{N,k,d}
{\cal O}_{e^{N-2-\gamma}}.
\end{equation}
We recall that ${\cal M}_{0,d,3}^{M_{N}^{k}}$ represents
the moduli space of rational
curves of degree $d$ with three punctures in $M_{N}^{k}$.
One should note that  ${\cal O}_{e}$ is
a multiplicative generator of $H^{*}_{q,e}(M_{N}^{k})$,
and therefore it is enough to  determine the multiplication rule
between ${\cal O}_{e}$ and ${\cal O}_{e^{\alpha}}$. The
topological selection rule yields that $C_{1\alpha\beta}^{N,k}$ is non-zero only if
$1+\alpha+\beta= N-2+(N-k)d$, hence it is:
\begin{equation}
{\cal O}_{e}\cdot {\cal O}_{e^{\alpha}}
= \sum_{d=0}^{\infty}q^{d}\frac{1}{k}C_{1,\alpha,
N-3-\alpha+(N-k)d}^{N,k,d}
{\cal O}_{e^{\alpha+1-(N-k)d}}.
\label{mul1}
\end{equation}
For conventional reason, we rewrite (\ref{mul1}) as follows:
\begin{eqnarray}
 {\cal O}_{e}\cdot {\cal O}_{e^{N-2-m}}
&=& {\cal O}_{e^{N-1-m}}+\sum_{d=1}^{\infty}q^{d}L_{m}^{N,k,d}
{\cal O}_{e^{N-1-m-(N-k)d}}\nonumber\\
L_{m}^{N,k,d}&:=&\frac{1}{k}C_{1,N-2-m, m-1+(N-k)d}^{N,k,d}
\label{mul2}
\end{eqnarray}
Here we have used the fact that
$q^{0}$ part of $H^{*}_{q,e}(M_{N}^{k})$ coincides with
the classical cohomology ring. The
integers $L_{m}^{N,k,d}$ are
the structural constants of the quantum ring. One
should think of  $k L_{m}^{N,k,d}$ as the number of rational
curves of degree $d$ on $M_{N}^{k}$ which meet a linear
section of dimension $m$ and a second linear section
of the right ( $ = m + (N-k)d -1  $) codimension.
We shall see that  $L_{m}^{N,k,1}$ are independent
of $N$ if  $N \geq k+2 $, and therefore we write them
simply as  $L_{m}^{k,1}$; we  refer to them
as the Schubert number of lines. Note that
 $L_{m}^{k,1} =   L_{k-m-1}^{k,1}$.

The preceding vanishing conditions translate into
\begin{eqnarray}
L_{m}^{N,k,d}\neq 0&\Longrightarrow& 0\leq m\leq (N-1)-(N-k)d
\quad (N-k\geq2)\nonumber\\
&\Longrightarrow&  1\leq m\leq (N-3)\quad(N-k=1, d=1)\nonumber\\
&\Longrightarrow&  0\leq m\leq (N-1)-(N-k)d\quad(N-k=1, d\geq 2)
\nonumber\\
&\Longrightarrow&  2\leq m\leq (N-3)\quad(N-k=0)
\label{flat}
\end{eqnarray}
We remark explicitly that if the dimension $N$ is large with respect to $k$
($N\geq 2k$) then
the only non trivial  quantum correction left
is due to curves of degree 1.

As we said above ${\cal O}_{e}$ is a multiplicative generator  of the ring,
and then there are coefficients $\gamma$ which give the representations:
\begin{equation}
{\cal O}_{e^{N-1-m}}= ({\cal O}_{e})^{N-1-m}-
                      \sum_{d=1}^{\infty}q^{d}\gamma_{m}^{N,k,d}
                        ({\cal O}_{e})^{N-1-m-(N-k)d}
\label{g}
\end{equation}
When we set $m=0$   we obtain the
main relation
of $H^{*}_{q,e}(M_{N}^{k})$.
        One has
\begin{eqnarray}
&&{\cal O}_{e}\cdot(({\cal O}_{e})^{N-2-m}-
                      \sum_{d=1}^{\infty}q^{d}\gamma_{m+1}^{N,k,d}
                        ({\cal O}_{e})^{N-2-m-(N-k)d})\nonumber\\
&=&  ({\cal O}_{e})^{N-1-m}-
                      \sum_{d=1}^{\infty}q^{d}\gamma_{m}^{N,k,d}
                        ({\cal O}_{e})^{N-1-m-(N-k)d}\nonumber\\
&+&\sum_{d=1}^{\infty}L_{m}^{N,k,d}q^{d}
(({\cal O}_{e})^{N-1-m-(N-k)d}-
\sum_{d'=1}^{\infty}q^{d'}\gamma_{m+(N-k)d}^{N,k,d'}
                        ({\cal O}_{e})^{N-1-m-(N-k)(d+d')})
\label{cons}
\end{eqnarray}
and therefore it is:
\begin{equation}
\gamma_{m}^{N,k,d}-\gamma_{m+1}^{N,k,d}=
L_{m}^{N,k,d}-\sum_{d'=1}^{d-1}L_{m}^{N,k,d-d'}
\gamma_{m+(N-k)(d-d')}^{N,k,d'}.
\label{recm}
\end{equation}
This yields:
\begin{equation}
\gamma_{m}^{N,k,d}=
\sum_{l=1}^{d}\sum_{\sum_{i=1}^{l}d_{i}=d}(-1)^{l-1}
\sum_{j_{l}=m}^{N-1-(N-k)d}\cdots
\sum_{j_{2}=m}^{j_{3}}\sum_{j_{1}=m}^{j_{2}}
(\prod_{i=1}^{l}L^{N,k,d_{i}}_{j_{i}+(\sum_{n=1}^{i-1}d_{n})(N-k)})
\label{transm}
\end{equation}
The fact that  the main relation of $H^{*}_{q,e}(M_{N}^{k})$ is of
the form of $({\cal O}_{e})^{N-1}-k^{k}({\cal O}_{e})^{k-1}\cdot q=0$
\cite{jin,givental}
is equivalent to
\begin{eqnarray}
\gamma_{0}^{N,k,1}&=&\sum_{j=1}^{N-1}L_{j}^{N,k,1}=k^{k}\nonumber\\
\gamma_{0}^{N,k,d}&=&
\sum_{l=1}^{d}\sum_{\sum_{i=1}^{l}d_{i}=d}(-1)^{l-1}
\sum_{j_{l}=0}^{N-1-(N-k)d}\cdots
\sum_{j_{2}=0}^{j_{3}}\sum_{j_{1}=0}^{j_{2}}
(\prod_{i=1}^{l}L^{N,k,d_{i}}_{j_{i}+(\sum_{n=1}^{i-1}d_{n})(N-k)})=0
\nonumber\\
&&\;\;\;\;\;\;\;\;\;\;\;\;(d\geq 2).
\label{reltr}
\end{eqnarray}
\subsection{The role of primitive cohomology}
In this subsection  we consider  the
general structure of the quantum cohomology
ring of a Fano hypersurface $V$
of degree $k$ in $ {\bf P}^{n+1}$ ($ n \geq 3 $) including
the primitive part.

It is
$ H_2 (V , {\bf Z} ) = {\bf Z} q  $, where
$kq$ is the class of a plane section, and
$ H^2 (V ,{\bf Z})$  is spanned by the
class $x (:=e)$ of the hyperplane section $H$.
The ring $ H^{\ast} (V , {\bf Q}) $ is generated
by $x$ and by the primitive cohomology
$ H^n (V , {\bf Q})_0 $, with the relations
$ x^{ n+1 } = 0 $, $ x \cup a_1 = 0 $,
$ a_1 \cup a_2 =  k^{-1}\int _V (a_1\wedge a_2 )x^n$
for  $ a_1$, $a_2$ primitive classes. We
shall denote $(\, | \,)_V$ the intersection
form, hence $(a|b)_V= \int _V a\wedge b$.
For $ 0 \leq i \leq n $, $x_{i}(:=e^{i})$ is
the class of the linear section of $V$
of codimension $i$, so that $x=x_1$. The vectors $x_{i}$
span the invariant part $R$ of $ H^{\ast} (V , {\bf Q}) $,
this is the orthogonal complement of $ H^n (V , {\bf Q})_0 $.
We recall that the Fano index
of V is $ h = h(V) = n+2 - k $.
Denote   ${\bf Z} \{ H_2 (V , {\bf Z}) \} $
the graded homogeneous
ring of formal series $ \sum n_d q^d $ with integer coefficients.
One introduces a ring structure on
$ H^{*}  (V , {\bf Z} \{ H_2 (V , {\bf Z}) \} )$ by the rule that
for homogeneous
$ \alpha^* $, $ \beta^* $ in $ H^{*}  (V , {\bf Z} ) $
the quantum multiplication product is
$ \alpha^*  \cdot \beta^* =
\sum_l (\alpha^* ,\beta^*)_d q^d$, where
$(\alpha^*,\beta^*)_0$ is the ordinary cohomology
product, and $(\alpha^*,\beta^*)_d$ is a class
of degree $ deg(\alpha^*) + deg(\beta^*) -2hd$ defined by
the condition $ ((\alpha^*,\beta^*)_d|\gamma) =
[\alpha^*,\beta^*,\gamma;d;V](= \langle {\cal O}_{\alpha^{*}}
{\cal O}_{\beta^{*}}
{\cal O}_{\gamma}\rangle_{V,d,gravity})$.
This last term is the GW invariant,
which can be informally defined as
the number of rational
curves of degree $d$ on $V$ meeting representative
submanifold $A$,$B$,$G$ in general position.
We shall use
the associativity and the
grading properties of $\cdot$, whose
rigorous and highly non trivial
construction is due to Ruan and Tian \cite {ruantian}.
We recall some facts from \cite {tian}.
Tian observed that the GW classes $[\alpha_1,\dots,\alpha_l;d;V]$
are invariant under monodromy action,
this is a direct corollary of
the main result in [RT], and he applied this explicitly to
cases like hypersurfaces by using the Picard-Lefschetz theorem.
\begin{prop}{\bf Tian}
If $m-l$ is odd and
$ a_s $ are  primitive classes then $$ [x_{i_1},\dots,x_{i_l},
a_{l+1},\dots,a_m;d;V] = 0 .$$
\end{prop}
{\it Proof}
The statement holds when $n$ is odd
for trivial reasons, indeed by definition\\
$[x_{i_1},\dots,x_{i_l},
a_{l+1},\dots,a_m;j;V] = 0$  if
$ 2(\sum i_j) + (m-l)n \neq 2n + 2hd + 2(m+l-3) $.
Coming to the case when the hypersurface $V$
is even dimensional, we recall that
the monodromy group $M$ is generated by reflections defined
by the vanishing cycles.
The case of even dimensional quadrics
is readily checked, since the vanishing cohomology
has rank one in this case. On the other hand
if $n > 3$  and $k>3$, by the same argument explained
in p.384 of \cite {PP}, a lemma of
Deligne yields that the Zariski closure
$ \bar M $ is in fact the full group of
isometries of $ H^{\ast} (V , {\bf C})_0 $.
Thus the GW invariant above
defines a symmetric multilinear form
with an odd number of entries,
invariant under the orthogonal group,
it is clear that such a form vanishes.

 If $ h \geq 2$ Tian's result yields
$ x \cdot a = 0 $, for $a \in H^n (V ,  {\bf Q} )_0$.
Instead we have
\begin{prop} If  $h = 1$
then $ x \cdot a = k!aq $.
\end{prop}

{\it Proof}
The statement is equivalent to
$ [x,a_1,a_2;1]  = - k!(a_1|a_2)_V $,
here $a_i$ are primitive classes and  $ [x,a_1,a_2;1] $
is the GW   number of the  lines which
meet them.
Our proof of this equality  is based
on  a remark of Beauville, \cite  {beauville}
$4.$ Application II. In this direction we also
need to prove the  formula below,
which is  a generalization
of a result of Tyurin, \cite {Ty}, \cite {blmur} and \cite {L}.
Let $W$ be a general hypersurface whose generic
hyperplane section is $V$. Then
the Fano variety $F(W)$ of
lines on $W$ is non singular irreducible
of dimension $ k $ and there
are $ k! $ lines on $W$ which meet a general point,
\cite {L} . The variety  $F(V)$
is a non singular  subvariety of
codimension $2$ in $F(W)$.
The natural $ {P}^1$
bundle $ p: L \rightarrow F(W)$
surjects  $ \lambda : L \rightarrow W$ with degree $k!$.
We denote  $\gamma:  BF \to V$ the restriction
of  $ \lambda $ to $V$, $\gamma$
has degree $k!$. Then  $\beta:  BF \to F(W)$ is
the blow up along $F(V)$ and the
projection of the exceptional
divisor  $\pi: E \to F(V)$ is the restriction of $p$.
We denote here  $ i: E \to  BF $ and $j: V \to W$
the natural inclusions.
The cohomology of
a blow up decomposes as a direct sum, in our case
$ H^*(BF, {\bf Q}) =
 i_{ \ast } \pi^{ \ast } (H^ {*-2}(F(V),{\bf Q}))
 \oplus  \beta ^{ \ast } (H^*(F(W),{\bf Q})) $.
Now  $\gamma_{\ast} H^*(BF,{\bf Q}) \to H^* (V,{\bf Q}) $
is a surjection,
because $ \gamma: BF \to V$ is.
It is known that the primitive cohomology
is contained in
the image  $ (\gamma i)_{\ast} \pi^{\ast}(H^ {*-2}(F(V),{\bf Q})) $,
\cite {L}. We need the stronger result that
given a primitive class $ a $ there is a class
$ \alpha $ with $ \gamma^{\ast} (a) =
i_\ast \pi^{\ast} \alpha $. To prove this statement
we first note that it is
equivalent to $   \beta_{\ast} \gamma^{\ast} (a) = 0 $,
and then we consider a  cycle $A$ which
represents $ a $ and which is
in general position with respect to the locus
covered by the lines on $V$.
We have $ \beta_{\ast} \gamma^{\ast} (A)$ $=$
$  \beta_{\ast} (\lambda ^{\ast}(j_{\ast}(A)) \cap BF) $,
and then  $ j_{\ast}(A) = 0 $ in $H^ {n+2}(W, \bf Q))$
because primitive classes are annihilated
by $ j_{\ast}$.
Fix next  $a_1$ and $a_2$ primitive classes
so that
$ \gamma^\ast a_1 =  i_\ast \pi^{\ast} \alpha_1 $,
$ \gamma^\ast a_2 =i_\ast \pi^{\ast} \alpha_2 $.
One has equality of degrees of intersection
$ (\gamma^\ast a_1|\gamma^\ast a_2)_{BF}
=k!(a_1|a_2)_V$, because
the degree
of $\gamma$ is $k!$. On the other hand
the excess intersection  formula of \cite {Fu1}
yields  $ (i_\ast \pi^{\ast} \alpha_1|i_\ast \pi^{\ast} \alpha_2)_{BF}$
$=$
$ -(\pi^{\ast} \alpha_1 | \pi^{\ast}\alpha_2 \cdot \zeta  )_E $
$=$
$ -(\alpha_1|\alpha_2)_{F(V)}$,
here $\zeta  $ denotes the tautological class of $E$
as a $ P^1$ bundle, and  $\zeta  $ is
known to be the opposite of
the class of the normal bundle of $E$
in $BF$. Thus
$$ k!(a_1|a_2)_V = -(\alpha_1|\alpha_2)_{F(V)} .$$
Now it is geometrically clear,
and this is the idea from \cite {beauville}, that
$$ [x,a_1,a_2;1]  =
(\pi _  \ast  i^\ast \gamma^\ast a_1 |
 \pi _   \ast i^ \ast \gamma^\ast a_2 )_ { F(V) } =
(\alpha_1|\alpha_2)_{F(V)}. $$

Tian's vanishing implies  also that
the quantum product of the hyperplane class
with a linear section is of type

$$  x \cdot x_{s-1}  = x_{s} +
 \sum_{i \geq 1}  a_{d,s} x_{s-dh} q^d, $$
where $a_{d,s} =
k^{-1} [ x_{s-1}, x_{ n +dh - s} , x ; d]$ are
the structural constants.

We set  $w:=x + k!q$, if
$ h= 1$, and otherwise $ w:= x$,
and we write $w^{s}$ the $s-th$
power of $w$ with respect to the
quantum product. Then
$w$ satisfies a unique minimal
monic equation $F=0$,
of degree $(n+1)$,
the equation which is found by setting
$s = n+1$ in the displayed formula.
This is of the
form $F:= w^{n+1} + \sum_{d=1} ^{ [(n+1)/h] } c_d w^{n+1-dh}q^d$.
For primitive classes $a$ and $b$ we have
$ a \cdot b = k^{-1}(a|b)_V (w^{n} + \sum_{d=1} ^{ [n/h] }
b_d w^{n-dh}q^d)$. Following Tian we note that associativity
yields $ 0 = (w \cdot a) \cdot b =
k^{-1} (a|b)_V(w^{n+1} + \sum_{d=1} ^{ [n/h] } b_d w^{n+1-dh}q^d)$,
and thus $c_d = b_d$ for $ 1 \leq d \leq n $,
$c_{n+1} = 0$. Beauville in \cite {beauville} studied the structure of
the quantum ring of Fano hypersurfaces of
degree small with respect to the dimension.
Beauville's result deals with the case
$n \geq 2k -3 $, in this case only the
coefficient $c_1 \neq 0$. Now $-c_1$
is the sum of the Schubert numbers of
lines on $V$, and it turns out
that $-c_1 = k^k$ hence:

\begin{theorem}
The quantum cohomology of $V$ over the
rational numbers is generated by $w$ and $ H^n (V , {\bf Q})_0 $
with relations
$(i) \, w^{n+1} = k^k w^{k-1}q$, $(ii) \,  w \cdot a = 0 $,
$(iii) \, a \cdot b = k^{-1} (a|b)_V(w^n  - k^kw^{k-2}q)$.
\end{theorem}

This theorem  holds in fact always, the hardest part (i)
is a deep theorem of Givental \cite{givental}, while (ii) and (iii)
follow
 from the same arguments used before.
\smallskip

\section{ Recursion relations for the structure constants
of Fano hypersurfaces.}

        This section is devoted to the proof of the following
recursion laws and of some related results:
\begin{theorem}
 Consider a hypersurface $M_{N}^{k}$ in $CP^{N-1}$
of degree $k$, if the
1st Chern class $N-k \geq 2$ then the basic
structure constants satisfy the following recursion
relations:
\begin{eqnarray}
L^{N,k,1}_{m}&=&L^{N+1,k,1}_{m}:=L^{k}_{m}\\
L^{N,k,2}_{m}&=&\frac{1}{2}(L^{N+1,k,2}_{m-1}+L^{N+1,k,2}_{m}
+2L^{N+1,k,1}_{m}\cdot L^{N+1,k,1}_{m+(N-k)})\\
L^{N,k,3}_{m}&=&\frac{1}{18}(4L^{N+1,k,3}_{m-2}+10L^{N+1,k,3}_{m-1}
+4L^{N+1,k,3}_{m}\nonumber\\
&&+12L^{N+1,k,2}_{m-1}\cdot L^{N+1,k,1}_{m+2(N-k)}
+9L^{N+1,k,2}_{m}\cdot L^{N+1,k,1}_{m+2(N-k)}\nonumber\\
&&+6L^{N+1,k,2}_{m}\cdot L^{N+1,k,1}_{m+1+2(N-k)}\nonumber\\
&&+6 L^{N+1,k,1}_{m-1}\cdot L^{N+1,k,2}_{m-1+(N-k)}
+9 L^{N+1,k,1}_{m}\cdot L^{N+1,k,2}_{m-1+(N-k)}\nonumber\\
&&+12 L^{N+1,k,1}_{m}\cdot L^{N+1,k,2}_{m+(N-k)}\nonumber\\
&&+18L^{N+1,k,1}_{m}\cdot L^{N+1,k,1}_{m+(N-k)}\cdot
L^{N+1,k,1}_{m+2(N-k)})
\label{rec0}
\end{eqnarray}
\label{req1}
\end {theorem}

                Our arguments are heuristic.
We embed $X :=M_{N}^{k}$ as the linear section of a general
hypersurface $ Y:= M_{N+1}^{k}$ in $CP^{N}$  so that
\begin{equation}
M_{N}^{k}=M_{N+1}^{k}\cap H
\label{sect}
\end{equation}
where the hyperplane $H$ is identified with $CP^{N-1}$.
Next we introduce the notation
\begin{equation}
\langle {\cal O}_{e^{a_{1}}}{\cal O}_{e^{a_{2}}}\cdots {\cal
O}_{e^{a_{m}}}
\rangle_{M_{N}^{k},d,gravity}= [A^{N}_{a_{1}},A^{N}_{a_{2}}, \cdots,
A^{N}_{a_{m}};d,N,k].
\label{lin}
\end{equation}
Here the spaces $A^{N}_{a_{i}}$ are linear subspaces
of codimension $a_{i}$ in $CP^{N-1}$ and
in general position, so that
\begin{equation}
PD_{M_{N}^{k}}(e^{a_{i}})= A^{N}_{a_{i}}\cap M_{N}^{k}.
\label{pd}
\end{equation}
We define below the ``special position'' correlation functions.
\begin{equation}
G[A^{N+1}_{a_{1}}\cap H ,A^{N+1}_{a_{2}}\cap H, \cdots,A^{N+1}_{a_{m}}
\cap H;d,N+1,k]
\label{sing}
\end{equation}
Clearly  $A^{N+1}_{a_{i}}\cap H$'s is
a linear subspace
in $CP^{N}$ of codimension $a_{i}+1$  which lies in $CP^{N-1}= H$.
The special position correlation
function should count the number of rational curves
of degree $d$ on $Y$ with $m$ labeled points
on them which belong to the corresponding
linear spaces and which have the further
property that points with different labels
stay distinct. By taking the linear spaces in general
position in $CP^{N-1}$ we may assume that
$[A^{N+1}_{a_{1}}\cap H ,A^{N+1}_{a_{2}}\cap H, \cdots,A^{N+1}_{a_{m}}
\cap H;d,N,k]$ has no contribution from reducible
curves on $X$. Now an irreducible curve of degree $d$ which cuts
$H$ in  $d+1$ points lies on it and then

\begin{eqnarray}
 &&[A^{N}_{a_{1}},A^{N}_{a_{2}}, \cdots,A^{N}_{a_{d+1}};d,N,k]
+ R
\nonumber\\
&=& G[A^{N+1}_{a_{1}}\cap H ,A^{N+1}_{a_{2}}\cap H, \cdots,
A^{N+1}_{a_{d+1}}\cap H;d,N+1,k]
\label{sep}
\end{eqnarray}
where $R$ measures the contributions due to the connected
reducible curves on $Y$ which satisfy the conditions.
In the cases the we consider $R$ does not occur
for lines and conics  and it is a finite set
for the case of cubic curves, as we compute below. For curves
of degree $4$ or more  the family
of reducible curves supporting $R$ may be of positive dimension
and we are not able to determine the contribution due to them.
For this reason we shall restrict to the case
of curves of degree $d$ at most equal to $3$.

The following lemma, the specialization formula,
gives a procedure for
computing the degree of $G[A^{N+1}_{a_{1}}\cap H ,A^{N+1}_{a_{2}}\cap H,
\cdots,
A^{N+1}_{a_{d+1}}\cap H;d,N+1,k]$,
because
by definition
$ G[A^{N+1}_{a_{1}+1},A^{N+1}_{a_{2}+1},
\cdots, A^{N+1}_{a_{m}+1};d,N+1,k] = $
$[A^{N+1}_{a_{1}+1},A^{N+1}_{a_{2}+1},
\cdots, A^{N+1}_{a_{m}+1};d,N+1,k] $. By moving
$A^{N+1}_{a_{s+1}+1}$ into $A^{N+1}_{a_{s+1}}\cap H$
one has

\begin{lem}

\begin{eqnarray}
&& G[A^{N+1}_{a_{1}}\cap H , \cdots,
A^{N+1}_{a_{s}}\cap H,
A^{N+1}_{a_{s+1}+1},A^{N+1}_{a_{s+2}+1},
\cdots, A^{N+1}_{a_{s+t}+1};d,N+1,k]\nonumber\\
&=& G[A^{N+1}_{a_{1}}\cap H , \cdots,
A^{N+1}_{a_{s}}\cap H,
A^{N+1}_{a_{s+1}}\cap H ,A^{N+1}_{a_{s+2}+1},\cdots,
A^{N+1}_{a_{s+t}+1};d,N+1,k]+\nonumber\\
&&\sum_{j=1}^{s} G[A^{N+1}_{a_{1}}\cap H ,
\cdots,A^{N+1}_{a_{j-1}}\cap H,A^{N+1}_{a_{j}+a_{s+1}}\cap H ,
A^{N+1}_{a_{j+1}}\cap H ,
\cdots,A^{N+1}_{a_{s}}\cap H,\nonumber\\
&& A^{N+1}_{a_{s+2}+1},\cdots,
A^{N+1}_{a_{s+t}+1};d,N+1,k]\nonumber\\
\label{spec}
\end{eqnarray}
\end{lem}

\smallskip

       Here we explain the definition of
the special position G-W invariants.

        Given a projective variety $ Z $
Kontsevich [K] has constructed  the coarse moduli space
$\bar M := \bar M(Z,m,\beta) $ of stable maps of homological
class  $ \beta $ to $Z$. $\bar M $ is the set
of equivalence classes of data $[C,p_1, \dots, p_m,\mu]$
where $ \mu: C \to Z $ is the 'stable' map,
$ C $ is a varying,projective,
connected, nodal curve of arithmetic genus $0$,
and $ p_1, \dots, p_m $ are
distinct, labeled
nonsingular points  on $C$.
We refer to [FP] for a detailed discussion
of this construction. The
canonical evaluation maps
$ \rho_i : \,\bar M \to  Z $ are defined by
$\rho_i([C,p_1, \dots, p_m,\mu]) = \mu(p_i)$.
The interior $ M(Z,m,\beta) $ is the locus
in  $ \bar M(Z,m,\beta)$ corresponding
to nonsingular irreducible
domain curves. We write $ \bar M(Z,A,\beta) $
when the index set of labels is a set $A$ instead of
$ [n] = \{ 1, \dots \ ,n \}  $. There are forgetful maps
$ \phi _B : \bar M(Z,A,\beta) \to \bar M(Z,A-B,\beta) $, defined
when $B$ is a subset of $A$.
Let now  $Z$ be a general non singular
Fano hypersurface of dimension $ n \geq 3 $
and  index $ h(Z) = n+2 - deg(Z) $. We consider
the case when $ \beta $ is the class of
a curve of degree $d$ and we assume that
$\bar M(Z,m,d) $ has the expected dimension
$ dimZ + d h(Z) + m - 3 $ and similarly for
the boundary components. We recall that such components
are associated with the choice of a partition
$ A \cup B $
of the set $[m]:= \{ 1, \dots , m \} $ and of
the choice of $d_1$ and $d_2 $ with
$ d = d_1 + d_2 $.
The boundary component $ \bar D(A,B;d_1, d_2 )$
is defined as the locus of moduli
points corresponding to reducible
domain curve $ C = C_1 \cup C_2 $,
where $ \mu _{ \ast } (C_i ) $ has degree
$ d_i $. Here the curve $C$ is
obtained by gluing at $  \bullet $ the curve
$ C_1 $ which has
on it points marked by the elements
in $ A $ and a further point, labeled by $ \bullet $.
and $ C_2 $, which has
on it points marked by the elements
in $ B $ and a further point, also labeled by $  \bullet $.
There is an identification
$ \bar D(A,B;d_1, d_2 ) =
 \bar M(Z, A \cup \{ \bullet \},d_1)
\times _Z \bar M(Z,B \cup \{ \bullet \},d_2) $.

        In what follows we take $ n+2 = N $ and define
$ T_{i}, \, i =  1,...,m $ to be  linear spaces
of codimension  $t_{i} \geq 1$
in $ P^{n+2} $ and
in general position  there. As before $Y$ is a Fano
hypersurface of degree $k$.
We assume that  $ \sum t_i $
is the expected dimension $ dimY + d h(Y) + m - 3 $ of
$ \bar M(Y,m,d) $. We define
$ [t_{1},...,t_{m};d , Y] $ to be
the degree of the zero cycle
$[T_{1},...,T_{m};d , Y] $, which we define to be
the intersection
product  of the cycles  $ \rho_i  ^{-1} (T_{i}) $.
Here we assume that those cycles  intersect
transversally in a finite number of points, each one
which is associated with an irreducible source
curve and with the property that
the corresponding map sends different
labeled points to different images.  By definition
$ [t_{1},...,t_{m};d, Y] $ is
one of the  GW invariants of $Y$,  it is
called  basic if $ m=3 $  and at least one of the
$t_{i}$ is $1$.
The GW invariants on $X$ are defined in a similar way,
by means of linear spaces $ S_{i} \subset P^{n+1}$.
We shall use the convention that $ S_{i}$ and
$ T_{i}$ are spaces of the same dimension,
so that $ S_{i}$ is obtained by moving $ T_{i}$
into $P^{n+1}$.
\smallskip

        Given linear spaces as above we write
$ G[S_{1},\dots,S_{s},T_{s+1},\dots,T_{s+t} ;d;Y] $ to represent
the open cycle in $ \bar M(Y,s+t,d)$  which can be
informally  described as
the set of rational curves of
degree $\delta$ on $Y$ with $s+t$ marked points
such that the images of the labeled
points $p_{j}$ belong to the space with the
same label,  and such that for $ j \leq s $ and $ i \leq s $
if $p_{j}$ and $p_{i}$ have the same image point
in $S_{j} \bigcap S_{i}$ then
this point is a double point for the image
curve. We shall use the notation that $s_i$ is the codimension
of $S_i$ in $P^{n+1}$ so that $s_i = t_i +1$, because
of our convention. The codimension of the preceding cycle is
$ \sum_j (s_j +1) + \sum_j t_j $.
Our aim is to compute the degree
of $ G[S_{1},\dots,S_{s},T_{s+1},\dots,T_{s+t} ;d;Y] $
when its expected dimension
is $0$. By abuse of notations we shall often use the same notation
to represent both a cycle of dimension $0$ and  the degree of the said
cycle.
We define $ \bar M ^0 (s)$ to be the complement in
$  \bar M := \bar M(Y,s+t,d)$ of the union of
the components of type
$ \bar D(A,B;d_1 ,d_2 )$
with $ d_2 = 0 $ and with
at least two elements of $B$ which
are $  \leq s $. Thus we have
$ \bar M ^0 (s+1) \subset \bar M ^0 (s) \subset \bar M $.
 The evaluation $ \rho_i $
restricts to $ \rho(s)_i ^0 $ on $ \bar M ^0(s)$.
Our definition is that
$ G[S_{1},...,S_{s},T_{s+1},...,T_{s+t};d;Y]$ is
the intersection product
of the cycles  $ (\rho(s)_j ^o ) ^{-1} (S_{j}) $,$ j= 1,...s $,
$ (\rho(s)_l ^o ) ^{-1} (T_{l})$, $l= s+1,..., s+t $.
 If the
codimensions $s_j$ and $t_l$ are fixed
the set of lists  $(S_{1},...,S_{s},T_{s+1},...,T_{s+t})$
is parameterized by a product of Grassmann manifolds,
hence it is an irreducible variety and then
there is an open dense
subset of it where the degree of
$G[S_{1},...,S_{s},T_{s+1},...,T_{s+t}]$
is maximum. We shall assume that our lists come from this
subset.

        We start by noting that
$ [S_{1},...,S_{d+1};d;X]$ and
$ G[S_{1},...,S_{d+1};d;Y]$ both have
the same expected dimension, which we take to be $0$.

\begin{prop}
If the given cycles are
zero dimensional then

$G[S_{1},...,S_{d+1};d;Y]$
$ =$
$ [S_{1},...,S_{d +1};d;X] + R$,

where $R$ is supported on the boundary locus of
$ \bar M(Y,m,d) $  and more precisely
on the locus corresponding
to reducible domain curves with reducible
image.
\end{prop}

In order to compute
the degree of
$ [S_{1},...,S_{d+1};d;X]$ we need
to verify that the dimension of the
preceding cycles is in fact $0$ and then
to compute their degrees. Now we have by assumption
that $[S_{1},...,S_{d+1};d;X] $
has the correct dimension $0$, so the dimension of
$G[S_{1},...,S_{d+1};d;Y]$ can fail to be also
$ = 0 $ only if $R$ fails. Of course
the dimension of $R$ can be detected
by looking at decomposable curves on $Y$, that is
to the behavior of rational curves of degree strictly less
than $d$.

\smallskip

As we have said above the degree of $G[S_{1},...,S_{d+1};d;Y]$
is determined by a reduction procedure,
which is performed by moving linear
spaces $T_{i}$ which are in general
position in $P^{n+2}$ to  spaces
$S_{i}$ of the
same dimension, which are contained in the hyperplane $P^{n+1}$
and in general position there.
Our main tool is next proposition,
we have only heuristic arguments
to support it

\begin{prop}
Provided that the
dimensions of the cycles below are $0$ as it is expected
then
$ degree G[S_{1},...,S_{s},T_{s+1},...,T_{s+t} ;d;Y]$ $=$

 $ degree G[S_{1},...,S_{s},S_{s+1},T_{s+2},...,T_{s+t} ;d;Y] $
$ + $

$ degree \sum \psi _{ i } ^{-1} G[S_{1},\dots,S_{i-1},S_{i,s+1}
,S_{i+1},\dots,S_{s},T_{s+2},\dots,T_{s+t} ;d;Y] $,

\smallskip
\noindent where $ S_{i,s+1} := S_i \cap S_{s+1} $ and where
$ \psi_ { i }  $ is the isomorphism
$ \bar D([s+t] - \{ i, s+1 \},\{ i, s+1 \};d ,0) \to
 \bar M(Y, ([s+t] - \{ i, s+1 \}) \cup \{ \bullet \},d ) $.

\end{prop}

\smallskip

The specialization lemma is just a restatement of this
proposition.

The following procedure gives the recursive formulas :\\

1. By iterative application of the specialization formula  we write
$G[A_{a}^{N+1}\cap H, A_{b}^{N+1}\cap H, A_{1}^{N+1,1}\cap H,
\cdots,A_{1}^{N+1,d-1}\cap H;d,N+1,k]$ in terms of the
standard correlation functions  of $Y = M_{N+1}^{k}$.

2. We decompose the standard correlation functions found
in step $1$ as polynomials in the basic G-W invariants
of $Y$, by which we mean the functions
$\langle {\cal O}_{e^{a}}{\cal O}_{e^{b}}{\cal O}_{e}
\rangle_{d,M_{N+1}^{k},gravity}$. This step is done by means of
the first reconstruction theorem of Kontsevich and Manin
or, equivalently, by the microscopic version of the DWVV
equations.

3. We compute the contribution of reducible curves
in  $G[A_{a}^{N+1}\cap H, A_{b}^{N+1}\cap H, A_{1}^{1,N+1}\cap H,
\cdots,A_{1}^{d-1,N+1}\cap H;d,N+1,k]$.

The 1st step gives next equalities, in writing them
we use the convention that
if number of insertion points
gets lower than 3 then we insert an hyperplane condition
and divide by the degree of the curve.
\begin{eqnarray}
&&G[A_{a}^{N+1}\cap H,A_{b}^{N+1}\cap H;1,N+1,k]\nonumber\\
&=&[A_{a+1}^{N+1},A_{b+1}^{N+1};1,N+1,k]-[A_{a+b+1}^{N+1};1,N+1,k]\nonumber\\
&&(a+b = N-3+(N-k))\nonumber\\
\nonumber\\
&&G[A_{a}^{N+1}\cap H, A_{b}^{N+1}\cap H, A_{1}^{N+1}\cap
H;2,N+1,k]\nonumber\\
&=& [A_{a+1}^{N+1},A_{b+1}^{N+1},A_{2}^{N+1};2,N+1,k]
-[A_{a+2}^{N+1},A_{b+1}^{N+1};2,N+1,k]\nonumber\\
&-&[A_{a+1}^{N+1},A_{b+2}^{N+1};2,N+1,k]\nonumber\\
&-&[A_{a+b+1}^{N+1},A_{2}^{N+1};2,N+1,k]+
2[A_{a+b+2}^{N+1};2,N+1,k]\nonumber\\
&&(a+b=  N-3+2(N-k))\nonumber\\
\nonumber\\
&&G[A_{a}^{N+1}\cap H, A_{b}^{N+1}\cap H, A_{1}^{N+1,1}\cap H,
A_{1}^{N+1,2}\cap H;3,N+1,k]
\nonumber\\
&=&[A_{a+1}^{N+1}, A_{b+1}^{N+1}, A_{2}^{N+1,1},A_{2}^{N+1,2}
;3,N+1,k]\nonumber\\
&-&2[A_{a+1}^{N+1}, A_{b+2}^{N+1}, A_{2}^{N+1};3,N+1,k]
-2[A_{a+2}^{N+1}, A_{b+1}^{N+1}, A_{2}^{N+1};3,N+1,k]\nonumber\\
&-&[A_{a+b+1}^{N+1}, A_{2}^{N+1,1},A_{2}^{N+1,2};3,N+1,k]
-[A_{a+1}^{N+1}, A_{b+1}^{N+1}, A_{3}^{N+1};3,N+1,k]\nonumber\\
&+&2[A_{a+1}^{N+1}, A_{b+3}^{N+1};3,N+1,k]
+2[A_{a+2}^{N+1}, A_{b+2}^{N+1};3,N+1,k]\nonumber\\
&+&2[A_{a+3}^{N+1}, A_{b+1}^{N+1};3,N+1,k]\nonumber\\
&+&4[A_{a+b+2}^{N+1}, A_{2}^{N+1};3,N+1,k]
+[A_{a+b+1}^{N+1}, A_{3}^{N+1};3,N+1,k]\nonumber\\
&-&6[A_{a+b+3}^{N+1};3,N+1,k]\nonumber\\
&&(a+b=  N-3+3(N-k)).
\label{oka}
\end{eqnarray}
We assume now that the Fano index of $X$ is at least $2$,
namely  $N-k\geq 2$, then $a+b+1$ is greater than
$N=dim(M_{N+1}^{k})+1$, and therefore
(\ref{oka}) is truncated in an obvious way.
At this point we recall the
definition
$[A_{a_{1}}^{N+1},A_{a_{1}}^{N+1},\cdots,A_{a_{m}}^{N+1};d,N+1,k]$
$=$ $\langle {\cal O}_{e^{a_{1}}}{\cal O}_{e^{a_{2}}}\cdots,
{\cal O}_{e^{a_{m}}}\rangle_{d,
M_{N+1}^{k},gr}$. In order to
proceed we need to express $\langle
{\cal O}_{e^{a}}{\cal O}_{e^{b}}{\cal O}_{e^{2}}\rangle_{d,
M_{N+1}^{k},gr}$ and $\langle
{\cal O}_{e^{a}}{\cal O}_{e_{b}}{\cal O}_{e^{3}}\rangle_{d,
M_{N+1}^{k},gr}$ in terms of  $\langle
{\cal O}_{e^{a}}{\cal O}_{e^{b}}{\cal O}_{e}\rangle_{d,
M_{N+1}^{k},gr}$.
Our tool is the first reconstruction
theorem of Kontsevich and Manin \cite {km} it yields
\begin{eqnarray}
\langle{\cal O}_{e^{a}}{\cal O}_{e^{b}}{\cal O}_{e^{m}}\rangle
\langle{\cal O}_{e^{N-1-m}}{\cal O}_{e}
{\cal O}_{e}\rangle
&=& \langle
{\cal O}_{e^{a}}{\cal O}_{e}{\cal O}_{e^{m}}\rangle
\langle{\cal O}_{e^{N-1-m}}{\cal O}_{e^{b}}
{\cal O}_{e}\rangle\nonumber\\
 \langle {\cal O}_{e^{a}}{\cal O}_{e^{b}}{\cal O}_{e^{m}}\rangle
\langle{\cal O}_{e^{N-1-m}}{\cal O}_{e^{2}}
{\cal O}_{e}\rangle
&=& \langle {\cal O}_{e^{a}}{\cal O}_{e^{2}}{\cal O}_{e^{m}}\rangle
\langle{\cal O}_{e^{N-1-m}}{\cal O}_{e^{b}}
{\cal O}_{e}\rangle\nonumber\\
 \langle {\cal O}_{e^{a}}{\cal O}_{e^{b}}{\cal O}_{e^{2}}{\cal
O}_{e^{m}}
\rangle
\langle{\cal O}_{e^{N-1-m}}{\cal O}_{e}
{\cal O}_{e}\rangle
&+&
\langle {\cal O}_{e^{a}}{\cal O}_{e^{b}}{\cal O}_{e^{m}}
\rangle
\langle{\cal O}_{e^{N-1-m}}{\cal O}_{e^{2}}{\cal O}_{e}
{\cal O}_{e}\rangle\nonumber\\
=
\langle {\cal O}_{e^{a}}{\cal O}_{e}{\cal O}_{e^{2}}{\cal
O}_{e^{m}}\rangle
\langle{\cal O}_{e^{N-1-m}}{\cal O}_{e^{b}}
{\cal O}_{e}\rangle
&+&
\langle {\cal O}_{e^{a}}{\cal O}_{e}{\cal O}_{e^{m}}\rangle
\langle{\cal O}_{e^{N-1-m}}{\cal O}_{e^{b}}{\cal O}_{e^{2}}
{\cal O}_{e}\rangle\nonumber\\
\label{km}
\end{eqnarray}
We  find in the end , if
the Fano index of $X$ is at least $2$,
\begin{eqnarray}
&&\frac{1}{k} G[A^{N+1}_{N-2-m}\cap H,A^{N+1}_{m-1+(N-k)}\cap H;1,N+1,k]
\nonumber\\
              &=&L^{N+1,k,1}_{m}:=L^{k}_{m}\\
&&\frac{1}{k}G[A_{N-2-m}^{N+1}\cap H,
A_{m-1+2(N-k)}^{N+1}\cap H, A_{1}^{N+1}\cap H;2,N+1,k]
\nonumber\\
              &=&\frac{1}{2}(L^{N+1,k,2}_{m-1}+L^{N+1,k,2}_{m}
+2L^{N+1,k,1}_{m}\cdot L^{N+1,k,1}_{m+(N-k)})\\
&&\frac{1}{k}G[A_{N-2-m}^{N+1}\cap H, A_{m-1+3(N-k)}^{N+1}\cap H,
A_{1}^{N+1,1}
\cap H,A_{1}^{N+1,2}\cap H
;3,N+1,k]\nonumber\\
              &=&\frac{1}{6}(4L^{N+1,k,3}_{m-2}+10L^{N+1,k,3}_{m-1}
+4L^{N+1,k,3}_{m}\nonumber\\
&&+12L^{N+1,k,2}_{m-1}\cdot L^{N+1,k,1}_{m+2(N-k)}
+12L^{N+1,k,2}_{m}\cdot L^{N+1,k,1}_{m+2(N-k)}\nonumber\\
&&+6L^{N+1,k,2}_{m}\cdot L^{N+1,k,1}_{m+1+2(N-k)}\nonumber\\
&&+6L^{N+1,k,2}_{m-1+(N-k)}\cdot L^{N+1,k,1}_{m-1}
+12L^{N+1,k,2}_{m-1+(N-k)}\cdot L^{N+1,k,1}_{m}\nonumber\\
&&+12L^{N+1,k,2}_{m+(N-k)}\cdot L^{N+1,k,1}_{m}\nonumber\\
&&+18L^{N+1,k,1}_{m}\cdot L^{N+1,k,1}_{m+(N-k)}\cdot
L^{N+1,k,1}_{m+2(N-k)})
\label{reck}
\end{eqnarray}
We make the  hypothesis that the Schubert
varieties of conics and lines on $Y$ and on $X$ which are
associated with the
given linear spaces $S_i$, $T_j$ and their intersections
have the right dimension. A count of dimensions
shows that for the cases of  degree 1 and degree 2
there is no contribution from reducible curves  and therefore

$G[A_{N-2-m}^{N+1}\cap H,A_{m-1+(N-k)}^{N+1}\cap H;1,N+1,k]/k
= L^{N,k,1}_{m}$

and

$G[A_{N-2-m}^{N+1}\cap H,A_{m-1+(N-k)}^{N+1}\cap H,A_{1}^{N+1}
\cap H;2,N+1,k]/k
= L^{N,k,2}_{m}$.

 For cubic curves there is a contribution due to reducible connected
curves which are made
of a line lying on $X$ and of a
conic lying on $Y$. There are two cases which occur, in one instance
the line meets $A_{N-2-m}^{N+1}\cap H$,
the conic meets the line and  $A_{m-1+3(N-k)}^{N+1}\cap H$.
In the other case the incidence conditions with the linear
spaces are reversed. In this way
\begin{eqnarray}
&&G[A_{N-2-m}^{N+1}\cap H, A_{m-1+3(N-k)}^{N+1}\cap H, A_{1}^{N+1,1}
\cap H,A_{1}^{N+1,2}\cap H
;3,N+1,k]/k \nonumber\\
&=&
3L^{N,k,3}_{m}
+R \frac{1}{k}
\nonumber\\
&=&
3L^{N,k,3}_{m}
+\frac{1}{2}L^{N+1,k,2}_{m}\cdot L^{N,k,1}_{m+2(N-k)}
+\frac{1}{2}L^{N+1,k,2}_{m-1+(N-k)}\cdot L^{N,k,1}_{m}
\label{separe}
\end{eqnarray}
Q.E.D

We come now to the case of
Fano index $1$. The same type of computations
as above yield
\begin{theorem}
If $X$ is a hypersurface of degree
$k$ in $CP^{k}$
the  recursion
relations for the basic invariants of conics and cubic curves
are the same as given in Theorem 1,
instead the numbers of lines
satisfy the law
\begin{equation}
 L^{k+1,k,1}_{m}=L^{k+2,k,1}_{m}-L^{k+2,k,1}_{0}=
L^{k+2,k,1}_{m}-k!
\end{equation}
\end {theorem}
{\em Proof}
In this case, $a+b+1$ $=$ $N-2+d$, from (\ref{oka}).
\begin{eqnarray}
&&[A_{a}^{N+1}\cap H,A_{b}^{N+1}\cap H;1,N+1,k]\nonumber\\
&=&[A_{a+1}^{N+1},A_{b+1}^{N+1};1,N+1,k]-[A_{N-1}^{N+1};1,N+1,k]\\
&&(a+b+1 = N-1)\nonumber
\label{mod1}
\end{eqnarray}

Now we can prove:

\begin{cor}
The main relation of the quantum ring of a Fano hypersurface
$M_{N}^{k}$ with index $N-k \geq 2$
is of the form
\begin{equation}
 ({\cal O}_{e})^{N-1}-k^{k}({\cal O}_{e})^{k-1}q=0
\end{equation}
up to $q^{3}$.
\end{cor}
{\em Proof}
The recursion relations of Theorem 2. do not change
$\gamma_{0}^{N,k,d}$, namely $\gamma_{0}^{N,k,d}=
\gamma_{0}^{N+1,k,d}$. If $N\geq 2k+1$, then
$\gamma_{0}^{N,k,d}=0 (d\geq 2)$, because of the
vanishing conditions due to the topological
selection rule.On the other hand $\gamma_{0}^{N,k,1}=k^{k}$,
from Schubert calculus cf. \cite{beauville}.
\begin{cor}
The main relation of the quantum cohomology
ring of a Fano hypersurface of index $1$ and dimension $k-1$
is of the form
\begin{equation}
 ({\cal O}_{e}+k!q)^{k}-k^{k}({\cal O}_{e}+k!q)^{k-1}q=0
\end{equation}
up to $q^{3}$.
\end{cor}
{\em Proof}

Consider the multiplication rule (\ref{mul2}):

\begin{eqnarray}
 {\cal O}_{e}\cdot {\cal O}_{e^{k-1-m}}
&:=& {\cal O}_{e^{k-m}}+qL_{m}^{k+1,k,1}
{\cal O}_{e^{k-1-m}}+
\sum_{d=2}^{\infty}q^{d}L_{m}^{k+1,k,d}
{\cal O}_{e^{k-m-d}}
\label{mul2d}
\end{eqnarray}
This gives:
\begin{eqnarray}
 ({\cal O}_{e}+k!q) \cdot {\cal O}_{e^{k-1-m}}
&=& {\cal O}_{e^{k-m}}+q(L_{m}^{k+1,k,1}+k!)
{\cal O}_{e^{k-1-m}}
+\sum_{d=2}^{\infty}q^{d}L_{m}^{k+1,k,d}
{\cal O}_{e^{k-m-d}}\nonumber\\
&=&{\cal O}_{e^{k-m}}+\sum_{d=1}^{\infty}q^{d}\tilde{L}_{m}^{k+1,k,d}
{\cal O}_{e^{k-m-d}}.
\label{mul2e}
\end{eqnarray}
Set now $F={\cal O}_{e}+k!q$, use $F$ as a
multiplicative generator and write
\begin{equation}
{\cal O}_{e^{k-m}}= (F)^{k-m}-
\sum_{d=1}^{\infty}q^{d}\tilde{\gamma}_{m}^{k+1,k,d}
(F)^{k-m-d}
\quad (m=0,1,\cdots,k-1)
\label{gen}
\end{equation}
A standard computation yields $\tilde{\gamma}_{0}^{k+1,k,d}={\gamma}_{0}^{k+2,k,d}$,
and we conclude by descending induction as in the proof of the preceding corollary
gives  the wished relation.

Our last result is also easily proved by descending induction on $N$
\begin{cor} If
$d \leq 3 $ and  $N-k \geq 1$)
the structure constants
$L_{m}^{N,k,d}$  can be written as a homogeneous
polynomial of degree $d$
in the structure constants of lines $L_{m}^{k}$
\end{cor}

\section{Recursive formulas in the Calabi-Yau case.}

        Here we try to understand how the recursive formulas
change when the hypersurface becomes of Calabi-Yau type, i.e. when
we deal with  $M_{k}^{k}$ of degree  $k$ in $CP^{k-1}$. In this situation
we can proceed as before for lines and conics. On the other
hand we cannot  use the same method for curves of degree $3$,
because it is difficult in this case to control
the contribution from  reducible curves . We
give instead a conjectural recursive formulas for cubics.  In the last part
of the section we explain the trend of thought which led us to the conjecture.

        We recall first that given a general point on $M_{k}^{k}$
there are no rational curves  meeting it, and therefore $ L^{k,k,d}_{0} = 0$.

\begin{theorem}
The recursive laws for lines and conics on a Calabi-Yau hypersurface
of degree $k$    are for $m \geq 1$
\begin{eqnarray}
 L^{k,k,1}_{m}&=&L^{k+1,k,1}_{m}-L^{k+1,k,1}_{1}\\
L^{k,k,2}_{m}&=&\frac{1}{2}(L^{k+1,k,2}_{m-1}+L^{k+1,k,2}_{m}
-L^{k+1,k,2}_{0}-L^{k+1,k,2}_{1}
\nonumber\\
&&+2(L^{k+1,k,1}_{m}-L^{k+1,k,1}_{1})^{2})
\end{eqnarray}
\end{theorem}
{\em Proof of Theorem 4}

We go back to the specialization formula (\ref{oka}),
which we use with the condition $a+b = N-3$
because now we have $N=k$. Repeated use of the first reconstruction
theorem yields
\begin{eqnarray}
&&\frac{1}{k}G[A_{k-2-m}^{k+1}\cap H,A_{m-1}^{k+1}\cap H;1,k+1,k]\nonumber\\
              &=&L^{k+1,k,1}_{m}-L^{k+1,k,1}_{1} \\
&&\frac{1}{k}G[A_{k-2-m}^{k+1}\cap H, A_{m-1}^{k+1}\cap H,
A_{1}^{k+1}\cap H;2,k+1,k]
\nonumber\\
              &=&\frac{1}{2}(L^{k+1,k,2}_{m-1}+L^{k+1,k,2}_{m}
-L^{k+1,k,2}_{0}-L^{k+1,k,2}_{1}\nonumber\\
&+&2L^{k+1,k,1}_{m}\cdot (L^{k+1,k,1}_{m}-L^{k+1,k,1}_{1}))
\label{rec}
\end{eqnarray}
Next we check the contribution from reducible curves.
For lines there are no reducible curves, so that
\begin{equation}
\frac{1}{k}G[A_{k-2-m}^{k+1}\cap H,A_{m-1}^{k+1}\cap H;1,k+1,k]=L_{m}^{k,k,1}.
\label{cy1}
\end{equation}
In case of conics, the reducible curves are given by two intersecting lines, one lying
on $M_{k}^{k}=M_{k+1}^{k}\cap H$ and the other
on $M_{k+1}^{k}$, hence it is
\begin{eqnarray}
\frac{1}{k}G[A_{k-2-m}^{k+1}\cap H, A_{m-1}^{k+1}\cap H,
A_{1}^{k+1}\cap H;2,k+1,k]
=L^{k,k,2}_{m}+L^{k+1,k,1}_{1}\cdot L^{k,k,1}_{m}
\label{conic}
\end{eqnarray}
Q.E.D.

        Next we deal with cubic curves, the Calabi-Yau condition
implies again  $ L^{k,k,3}_{m} =0 $ for $ 0 \leq  m \leq 1$,
our proposal for larger $m$ is the following:

\begin{conj}
The recursive law for curves of degree $3$ on $M_{k}^{k}$
and for $m \geq 2$ becomes:
\begin{eqnarray}
L^{k,k,3}_{m}&=&\frac{1}{18}(4L^{k+1,k,3}_{m-2}+10L^{k+1,k,3}_{m-1}
+4L^{k+1,k,3}_{m}\nonumber\\
&&-10L^{k+1,k,3}_{0}
-4L^{k+1,k,3}_{1}\nonumber\\
&&+12L^{k+1,k,2}_{m-1}\cdot L^{k+1,k,1}_{m}
+12L^{k+1,k,2}_{m}\cdot L^{k+1,k,1}_{m}\nonumber\\
&&+6L^{k+1,k,2}_{m}\cdot L^{k+1,k,1}_{m+1}\nonumber\\
&&+6L^{k+1,k,2}_{m-1}\cdot L^{k+1,k,1}_{m-1}
+12L^{k+1,k,2}_{m-1}\cdot L^{k+1,k,1}_{m}\nonumber\\
&&+12L^{k+1,k,2}_{m}\cdot L^{k+1,k,1}_{m}\nonumber\\
&&+18(L^{k+1,k,1}_{m}-L^{k+1,k,1}_{1})^{2}\cdot
(L^{k+1,k,1}_{m}+2 L^{k+1,k,1}_{1}))\nonumber\\
&&-\frac{1}{6}L^{k+1,k,2}_{m-1}\cdot L^{k+1,k,1}_{m}-
\frac{1}{6}L^{k+1,k,2}_{m}\cdot
L^{k+1,k,1}_{m}\nonumber\\
&&-\frac{3}{4}L^{k+1,k,2}_{0}\cdot L^{k+1,k,1}_{m}
-\frac{3}{4}L^{k+1,k,2}_{1}\cdot L^{k+1,k,1}_{m}  \nonumber\\
&&-\frac{5}{12}L^{k+1,k,2}_{1}\cdot L^{k+1,k,1}_{1}
 -\frac{5}{12}L^{k+1,k,2}_{0}\cdot L^{k+1,k,1}_{1} \nonumber\\
&&-\frac{1}{3}L^{k+1,k,2}_{1}\cdot L^{k+1,k,1}_{2}\nonumber\\
&&-3 L^{k+1,k,1}_{1}\cdot\frac{1}{2}(L^{k+1,k,2}_{m-1}+L^{k+1,k,2}_{m}
-L^{k+1,k,2}_{0}-L^{k+1,k,2}_{1}\nonumber\\
&&+2(L^{k+1,k,1}_{m}-L^{k+1,k,1}_{1})^{2}).
\label{rec3}
\end{eqnarray}
\end{conj}

We came to this formula by means of the following considerations.
Using Theorem 4  one  can compute
the main relation
of $H^{*}_{q,e}(M_{k}^{k})$ up to degree 2, this reads:
\begin{eqnarray}
&&(1-(k^{k}-(k-2)L_{1}^{k}-2L_{0}^{k})q\nonumber\\
&&-(2k^{k}L_{0}^{k}+(k-3)k^{k}L_{1}^{k}-3(L_{0}^{k})^{2}-
(2k-6)L_{1}^{k}L_{0}^{k}-\frac{(k-3)(k-2)}{2}(L_{1}^{k})^{2}\nonumber\\
&&-\frac{k}{2}L_{0}^{k+1,k,2}-\frac{k-2}{2}L_{1}^{k+1,k,2})q^{2}-\cdots)
({\cal O}_{e})^{k-1}=0
\label{cyr}
\end{eqnarray}

On the other hand one has from \cite{mnmj} and  \cite{jin}
that the main relation
can be written using the $k-2$ point correlation function
of the pure matter theory in the form
\begin{eqnarray}
&&\frac{k}{\langle \prod_{j=1}^{k-2}{\cal O}_{e}(z_{j})
\rangle_{M_{k}^{k},matter}}
({\cal O}_{e})^{k-1}=0\nonumber\\
\end{eqnarray}

and that it is

\begin{eqnarray}
\langle\prod_{j=1}^{k-2}{\cal O}_{e}(z_{j})\rangle_{M_{k}^{k},matter}
& =& k + k^{k+1}(1-2a_1 - (k-2) (b_1))q\nonumber\\
&+&  k^{2k+1}(1-2a_1-b_1 +3(a_1)^2 -2a_2 +2a_1\cdot b_1 \nonumber\\
&+& (k-2)(-b_1+4a_1\cdot b_1+2(b_1)^2-2b_2)\nonumber\\
&+&\frac{(k-2)(k-3)}{2}(b_1)^2)q^{2}+\cdots
\label{nontri}
\end{eqnarray}
Here
$$a_d  = \frac{(kd)!}{(d!)^{k}k^{kd}} \; ,\;\;\;\;
 b_d = a_d(\sum_{i=1}^{d}\sum_{m=1}^{k-1}\frac{m}{i(ki-m)})$$
are the coefficients of the hypergeometric
series associated to the solutions
$$W_{0}(x) = \sum_{d=0}^{\infty}a_{d}e^{dx},\;\;\;\;
W_{1}(x) = \sum_{d=1}^{\infty}b_{d}e^{dx}+W_{0}(x)x $$
of the differential equation
\begin{eqnarray}
&&((\frac{d}{dx})^{k-1}-e^{x}(\frac{d}{dx}+\frac{1}{k})
(\frac{d}{dx}+\frac{2}{k})\cdots(\frac{d}{dx}+\frac{k-1}{k}))W_{i}(x)=0
\nonumber\\
\label{diff}
\end{eqnarray}
By comparison of (\ref{cyr}) with (\ref{nontri}), we  notice  the following
equalities:
\begin{eqnarray}
k^{k}a_{1}&=&L_{0}^{k}\nonumber\\
k^{k}b_{1}&=&L_{1}^{k}\nonumber\\
k^{2k}a_{2}&=&\frac{1}{2}(L_{0}^{k+1,k,2}+2(L_{0}^{k})^{2})\nonumber\\
k^{2k}b_{2}&=&\frac{1}{4}(L_{1}^{k+1,k,2}+L_{0}^{k+1,k,2}
+2L_{1}^{k}L_{1}^{k})+L_{1}^{k}L_{0}^{k}.
\label{corre}
\end{eqnarray}

        These equalities can be organized more systematically by means
of generating functions. To this aim we need :
\begin{defi}
For arbitrary $N$ and $k$ let $\tilde{L}_{m}^{N,k,d}$ be the integer obtained by applying
formally the recursive laws of Theorem 2.
\end{defi}

{\em Remark.} One should note:
(i) $\tilde{L}_{i}^{k,k}(e^{x})=\tilde{L}_{k-1-i}^{k,k}(e^{x})$,
(ii) if the index $N - k$
is at least 2 then $\tilde{L}_{m}^{N,k,d}$ must be the
ordinary structural constant
$L_{m}^{N,k,d}$ of the Fano hypersurface.

Now we can rewrite (\ref{corre}) as
\begin{eqnarray}
k^{k}a_{1}&=&\tilde{L}_{0}^{k,k,1}\nonumber\\
k^{k}b_{1}&=&\tilde{L}_{1}^{k,k,1}\nonumber\\
k^{2k}a_{2}&=&\tilde{L}_{0}^{k,k,2}\nonumber\\
k^{2k}b_{2}&=&\frac{1}{2}\tilde{L}_{1}^{k,k,2}+
\tilde{L}_{1}^{k,k,1}\cdot\tilde{L}_{0}^{k,k,1}
\label{rew}
\end{eqnarray}
After having performed some numerical computations,
we  have noticed that also the
following  relations should hold true:
\begin{eqnarray}
k^{3k}a_{3}&=&\tilde{L}_{0}^{k,k,3}\nonumber\\
k^{3k}b_{3}&=&\frac{1}{3}\tilde{L}_{1}^{k,k,3}+
\frac{1}{2}\tilde{L}_{1}^{k,k,2}\cdot\tilde{L}_{0}^{k,k,1}
+\tilde{L}_{1}^{k,k,1}\cdot\tilde{L}_{0}^{k,k,2}.
\end{eqnarray}

Consider next  the generating functions:
\begin{eqnarray}
\tilde{L}_{i}^{k,k}(\tilde{q})
:= 1+\sum_{d=1}^{\infty}\tilde{L}_{i}^{k,k,d}\tilde{q}^{d}
\;,\;\;\;\;\tilde{q} :=   e^{x}
\label{gen0}
\end{eqnarray}
and define
\begin{eqnarray}
t:= x+ (\sum_{j=1}^{\infty}b_{j}k^{kj}e^{jx})/
(\sum_{j=0}^{\infty}a_{j}k^{kj}e^{jx}).
\label{period1}
\end{eqnarray}
The preceding equalities motivate us to expect:
\begin{eqnarray}
\tilde{L}_{0}^{k,k}(\tilde{q})=\sum_{j=0}^{\infty}a_{j}k^{kj}e^{jx},
\quad \tilde{L}_{1}^{k,k}(e^{x})= \frac{dt}{dx}.
\label{period2}
\end{eqnarray}

We use the virtual structural constants $\tilde{L}$ to define
a virtual quantum product determined by the action of a ring generator $G$
which operates according to the rules:
\begin{eqnarray}
G\cdot {\cal O}_{{e}^{m-1}}&=&\tilde{L}_{k-1-m}^{k,k}(e^{x}){\cal O}_{e^{m}}
\\
&&(m= 1,2,\cdots, k-2)\nonumber\\
G\cdot {\cal O}_{{e}^{k-2}}&=& 0  \quad .
\end{eqnarray}
We note the relation
$G = G\cdot 1 = \tilde{L}_{k-2}^{k,k}(e^{x}){\cal O}_{e}$.
We expect that the structure constants of the virtual action satisfy
the following equality, which in fact may be checked up to  $\tilde{q}^{3}$  using
the recursive laws for the Fano case:
\begin{eqnarray}
\prod_{i=0}^{k-1}\tilde{L}_{i}^{k,k}(e^{x})&=&
(1-k^{k}e^{x})^{-1}
\label{prop1}
\end{eqnarray}
and this yields the relation:
\begin{eqnarray} (1-k^{k}e^{x})\cdot(\tilde{L}_{0}^{k,k}
(e^{x}))^{2}\cdot (G)^{k-1}= 0
\label{prop2}
\end{eqnarray}

On the other hand  the {\it true} quantum cohomology ring satisfies a similar
multiplication rule:
\begin{eqnarray}
{\cal O}_{e}\cdot 1&=&{\cal O}_{e}\nonumber\\
{\cal O}_{e}\cdot {\cal O}_{e^{m-1}}&=&L_{k-1-m}^{k,k}(e^t){\cal O}_{e^{m}}
\quad(m=2,3,\cdots,k-2)\nonumber\\
{\cal O}_{e}\cdot {\cal O}_{e^{k-3}}&=&{\cal O}_{e^{k-2}}\nonumber\\
{\cal O}_{e}\cdot {\cal O}_{e^{k-2}}&=&0\\
\mbox{where}&&\nonumber\\
L_{i}^{k,k}(e^t)&:=& 1+\sum_{d=1}^{\infty}L_{i}^{k,k,d}e^{dt}.
\label{reco}
\end{eqnarray}
We can compute $L_{i}^{k,k}(e^{t})$ in concrete examples using
the method of torus localization
see  \cite{j} for details and results.
\smallbreak
Now we search for a transformation rule to pass from the virtual
to the true quantum multiplication. To this aim we find useful to introduce a formal
definition:

\begin{defi}
The commutative product $(*)$ between  differential operators
of the form $f(x)\frac{d^{m}}{dx^{m}}$
is given by:
\begin{equation}
(f(x)\frac{d^{m}}{dx^{m}})*(g(x)\frac{d^{n}}{dx^{n}})
=(f(x)\cdot g(x))\frac{d^{m+n}}{dx^{m+n}}.
\label{commu}
\end{equation}
Given the coordinate change $x=x(t)$ we define
a map from $\frac {d }{dx }$ operators to
$\frac {d }{dt }$ operators by means of the rule
\begin{equation}
f(x)\frac{d^{m}}{dx^{m}} \to f(x(t))(\frac{dt}{dx})^{m}\frac{d^{m}}{dt^{m}}.
\label{comtrans}
\end{equation}

\end{defi}

At this point we can relate the quantum products laws
using as an intermediate step
the product of differential operators. To start we
propose the correspondence
\begin{equation}
{\cal O}_{e}=\frac{d}{dt},\quad G=\frac{d}{dx}.
\label{differ}
\end{equation}
Then one has
\begin{eqnarray}
\frac{d}{dx}*{\cal O}_{{e}^{m-1}}&=
&\tilde{L}_{k-1-m}^{k,k}(e^{x}) {\cal O}_{e^{m}},
\;\;(m = 1,2,\cdots, k-2)\nonumber\\
\frac{d}{dx}*{\cal O}_{{e}^{k-2}}&=& 0,
\label{G}
\end{eqnarray}
and
\begin{eqnarray}
\frac{d}{dt}*1&=&{\cal O}_{e}\nonumber\\
\frac{d}{dt}*{\cal O}_{e^{m-1}}&=& L_{k-1-m}^{k,k}(e^t){\cal O}_{e^{m}}
\quad(m=2,3,\cdots,k-2)\nonumber\\
\frac{d}{dt}*{\cal O}_{e^{k-3}}&=&{\cal O}_{e^{k-2}},\quad
\frac{d}{dt}*{\cal O}_{e^{k-2}}=0.
\label{real}
\end{eqnarray}
It follows from (\ref{G}):
\begin{equation}
\frac{d}{dx}*1= \tilde{L}_{k-2}^{k,k}(e^{x})\cdot{\cal O}_{e}
=  \tilde{L}_{k-2}^{k,k}(e^{x})\cdot\frac{d}{dt}
= \frac{dt}{dx}\cdot\frac{d}{dt}
\label{ques}
\end{equation}
This equality suggests that (\ref{G}) and (\ref{real}) become isomorphic
if we use the transformation of differential operators defined above.
Compare now the coefficients for ${\cal O}_{e^{\alpha}}$ in (\ref{G}) with (\ref{real}),
then the wished isomorphism yields the equality
\begin{eqnarray}
\prod_{j=1}^{\alpha} ( \tilde{L}_{k-1-j}^{k,k}(e^{x(t)})
 \frac{dx}{dt} )
= \prod_{j=1}^{\alpha} L_{k-1-j}^{k,k}(e^{t})  .
\label{recur}
\end{eqnarray}
We  find in this way the transformation laws that we were looking for, they are:
\begin{equation}
\frac{\tilde{L}_{i}^{k,k}(e^{x(t)})}{\tilde{L}_{1}^{k,k}(e^{x(t)})}=
\tilde{L}_{i}^{k,k}(e^{x(t)}){\frac{dx}{dt}}
=L_{i}^{k,k}(e^{t}).
\label{tra}
\end{equation}

This is the rule that provides the recursive
formulas for curves of arbitrary degree $d$ on the Calabi-Yau
hypersurface $M_{k}^{k}$ once that we know the
recursive formulas for curves up to degree  $d$ on
Fano hypersurfaces. At this point we obtain the recursive formulas
for cubics in Conjecture 1 by means of elementary calculations.

{\bf Example}\qquad The true quantum cohomology ring of
the quintic Calabi-Yau threefold is:
\begin{eqnarray}
{\cal O}_{e}\cdot 1&=&{\cal O}_{e}\nonumber\\
{\cal O}_{e}\cdot {\cal O}_{e}&=&{\cal O}_{e^{2}}(1+575e^{t}+975375e^{2t}+
1712915000e^{3t}+\cdots )\nonumber\\
{\cal O}_{e}\cdot {\cal O}_{e^{2}}&=&{\cal O}_{e^{3}}\nonumber\\
{\cal O}_{e}\cdot {\cal O}_{e^{3}}&=&0.
\end{eqnarray}
while the associated virtual ring is:
\begin{eqnarray}
G\cdot 1&=&{\cal O}_{e}(1+770e^{x}+1435650e^{2x}+3225308000e^{3x}
+\cdots )\nonumber\\
G\cdot {\cal O}_{e}&=&{\cal O}_{e^{2}}(1+1345e^{x}+3296525e^{2x}+
   8940963625e^{3x}+\cdots )\nonumber\\
G\cdot {\cal O}_{e^{2}}&=&{\cal O}_{e^{3}}(1+770e^{x}+1435650e^{2x}
+3225308000e^{3x}+\cdots)\nonumber\\
G\cdot {\cal O}_{e^{3}}&=&0
\end{eqnarray}
Using   (\ref{tra}) we find that:
\begin{eqnarray}
575&=&1345-770\\
975375&=&3296525-1435650+770\cdot 770\\
&&-1345\cdot 770-770\cdot (1345-770)\nonumber\\
1712915000&=&8940963625-3225308000\nonumber\\
&&+2\cdot 770\cdot 1435650-770^3+1345\cdot (770)^2\nonumber\\
             &&-1345\cdot 1435650-3296525\cdot 770\nonumber\\
         &&-2\cdot 770\cdot (3296525-1435650+770\cdot 770-1345\cdot
           770)\nonumber\\
          &&+(\frac{3}{2}\cdot (770)^2-\frac{1}{2}\cdot 1435650)
\cdot(1345-770).
\end{eqnarray}
\section{On the construction of the recursive formulas for rational curves
of larger degree.}
Motivated by the preceding results,
we begin this section by proposing some conjectures. They are 
strong enough to allow in principle the
construction of the expected recursive formulas for curves of higher
degree, and we explicitly produce the law for $d=4,5$.
\begin{conj}
There are universal recursive polynomial laws which express
the structure constants $L_{m}^{N,k,d}$ on a Fano variety in terms of $L_{m}^{N+1,k,n}$  $(1\leq n
\leq d )$.
 The formulas have the following properties.

1. The form of the recursive polynomials is invariant if the index $N-k\geq2$, and
the equality $\gamma_{0}^{N,k,d}=\gamma_{0}^{N+1,k,d}$
for the coefficients in the fundamental relations
is a consequence of them.

2. If $N-k=1$ the recursive formulas change
only for the case of lines, i.e. $d=1$.
\end{conj}

We keep the notations of section $4$, so that 
$\tilde{L}_{m}^{N,k,d}$ represents the result
of a formal iteration of the recursive functions
for fixed $k$ down to any chosen $N$, and then
$\tilde{L}_{m}^{N,k}$ is the associated generating function.

\begin{conj}
Formal iteration of the laws of Conjecture 2 for descending $N$
down to the case $N=k$  yields the coefficients of the hypergeometric series used in
the mirror calculation,
i.e., it should be
\begin{eqnarray}
k^{dk}a_{d}&=&\tilde{L}_{0}^{k,k,d}\nonumber\\
k^{dk}b_{d}&=&\frac{1}{d}\tilde{L}_{1}^{k,k,d}+
\sum_{m=1}^{d-1}\frac{1}{m}\tilde{L}_{1}^{k,k,m}\cdot\tilde{L}_{0}^{k,k,d-m},
\end{eqnarray}
and the same  procedure gives the structure constants
of the quantum cohomology ring of the Calabi-Yau hypersurface
of degree $k$,  using the rule
\begin{equation}
L_{i}^{k,k}(e^{t}) = \frac{\tilde{L}_{i}^{k,k}(e^{x(t)})}
{\tilde{L}_{1}^{k,k}(e^{x(t)})}.
\label{contr}
\end{equation}
\end{conj}

{\bf Remark. } It is an immediate consequence of
the conjectures and of the vanishing conditions
of section $2$ that the structural constant $L_{m}^{N,k,d}$ is a polynomial
of degree $d$ in the constants $L_{m}^{k}$,  $N \geq k$ .

We proceed now to the construction of the recursive formulas
for the case $d=4$. Our method is based on the expectation that the specialization
procedure gives if not the right coefficients at least the right
monomials which appear in the recursive laws.
We formalize this below with a conjecture.

\smallbreak

We start by constructing  some
technical formulas 
for the factorization of the Gromov-Witten invariants.

Let $\{n_{*}\}:=\{n_{1},n_{2},\cdots,n_{l}\}$ and $ind(\{n_{*}\})
= \sum_{j=1}^{l}(n_{j}-1)$. We formally define $ind(\{\emptyset\})$
to be $0$. We have a formula for the correlation
functions (Gromov-Witten invariants) of the topological sigma model
on $M_{N+1}^{k}$ coupled to gravity, which reads:
\begin{eqnarray}
&& k^{-1} \langle    \cdot {\cal O}_{e^{a}} {\cal O}_{e^{n_{1}}} {\cal O}_{e^{n_{2}}}
\cdots  {\cal O}_{e^{n_{l}}} {\cal O}_{e^{b}}\rangle_{d,M_{N+1}^{k},gr} \nonumber\\
&=&\sum_{d_{1}=0}^{d}\sum_{d_{2}=0}^{d_{1}}\cdots
\sum^{d_{ind(\{n_{*}\})-1}}_{d_{ind(\{n_{*}\})}=0}
C^{d}(\{n_{*}\};d_{1}, d_{2}, \cdots,d_{ind(\{n_{*}\})})
\prod_{i=0}^{ind(\{n_{*}\})}
L_{n+1-a-i+(N-k+1)(d-d_{i})}^{N+1,k,d_{i}-d_{i+1}}\nonumber\\
&&\\
&&d_{0}:= d,\quad N-k+1\geq ind(\{n_{*}\})+1
\end{eqnarray}

The coefficients $ C^{d}(\{n_{*}\};d_{1},\cdots,d_{ind(\{n_{*}\})})$
which appear here
have the following properties:
\begin{eqnarray}
&& C^{d}(\{m\};d_{1},\cdots,d_{m-1})=1\\
&&C^{d}(\{n_{*}\}\cup\{1\};
d_{1},\cdots,d_{ind(\{n_{*}\})})=
dC^{d}(\{n_{*}\};d_{1},\cdots,d_{ind(\{n_{*}\})})
\end{eqnarray}
One can determine  $ C^{d}(\{n_{*}\};d_{1},\cdots,d_{ind(\{n_{*}\})})$
by means of the recursive relation,
\begin{eqnarray}
&& C^{d}(\{n_{*}\};d_{1},\cdots,d_{ind(\{n_{*}\})})\nonumber\\
&=&\sum_{\stackrel{\{l_{*}\}\amalg\{m_{*}\}=\{n_{*}\}/\{n_{l}\}}
{\{m_{*}\}\neq \emptyset}} (C^{d-d_{ind(\{l_{*}\})+n_{l}-1}}
(\{l_{*}\}\cup\{n_{l}-1\};d_{1}-d_{ind(\{l_{*}\})+n_{l}-1},\nonumber\\
&&\cdots,
d_{ind(\{l_{*}\})+n_{l}-2}-d_{ind(\{l_{*}\})+n_{l}-1})\cdot\nonumber\\
&& C^{d_{ind(\{l_{*}\})+n_{l}-1}}
(\{m_{*}\};d_{1+ind(\{l_{*}\})+n_{l}-1},\cdots,
d_{ind(\{n_{*}\})})\cdot d_{ind(\{l_{*}\})+n_{l}-1})
\nonumber\\
&+& C^{d-d_{ind(\{n_{*}\})}}
(\{n_{*}\}/\{n_{l}\}\cup\{n_{l}-1\} ;d_{1}-d_{ind(\{n_{*}\})},\cdots,
d_{ind(\{n_{*}\})-1}-d_{ind(\{n_{*}\})}).\nonumber\\
\label{facto}
\end{eqnarray}
{\it Proof}

We prove (\ref{facto}) by induction of $ind(\{n_{*}\})$. We denote
${\cal O}_{e^{n_{1}}}{\cal O}_{e^{n_{2}}}\cdots {\cal O}_{e^{n_{l}}}$ as
${\cal O}_{e^{\{n_{*}\}}}$ for brevity.
The first reconstruction theorem of KM  yields:
\begin{eqnarray}
\sum_{\stackrel{{\{l_{*}\}\amalg\{m_{*}\}}}{=\{n_{*}\}/\{n_{l}\}}}
\sum_{d_{0}=0}^{d}&& k^{-1} \langle{\cal O}_{e^{a}}{\cal O}_{e^{\{l_{*}\}}}
{\cal O}_{e^{b}}{\cal O}_{e^{n_{l}+ind(\{m_{*}\})-(N-k+1)d_{0}}}
\rangle_{d-d_{0},gr}\cdot\nonumber\\
&&\cdot k^{-1}
\langle{\cal O}_{e^{ind(\{l_{*}\})-(N-k+1)(d-d_{0})+
a+b}}{\cal O}_{e^{\{m_{*}\}}}{\cal O}_{e^{n_{l}-1}}{\cal O}_{e}\rangle_{d_{0},
gr}\nonumber\\
=\sum_{\stackrel{{\{l_{*}\}\amalg\{m_{*}\}}}{=\{n_{*}\}/\{n_{l}\}}}
\sum_{d_{0}=0}^{d}&&k^{-1}\langle{\cal O}_{e^{a}}{\cal O}_{e^{\{l_{*}\}}}
{\cal O}_{e^{n_{l}-1}}{\cal O}_{e^{b+1+ind(\{m_{*}\})-(N-k+1)d_{0}}}
\rangle_{d-d_{0},gr}\cdot\nonumber\\
&&\cdot k^{-1}
\langle{\cal O}_{e^{a-1+n_{l}+ind(\{l_{*}\})-(N-k+1)(d-d_{0})
}}{\cal O}_{e^{\{m_{*}\}}}{\cal O}_{e}{\cal O}_{e^{b}}\rangle_{d_{0},
gr}.\nonumber\\
\label{funi}
\end{eqnarray}
The l.h.s. of (\ref{funi}) has the contribution of $d_{0}=0$ and
$\{m_{*}\}=\{\emptyset\}$ because $ind(\{n_{*}\})-(N-k+1)d_{0}\leq -1$
and because the classical correlation function remains
non-zero only if the number
of operator insertion point equals $3$. Then we have
\begin{eqnarray}
\mbox{(the l.h.s.)  of (\ref{funi})}&=& k^{-1}\langle
{\cal O}_{e^{a}}
{\cal O}_{e^{\{n_{*}\}}}{\cal O}_{e^{b}}\rangle_{d,gr}.
\label{lhs}
\end{eqnarray}
On the other hand, we can rewrite the r.h.s. of (\ref{funi}) from the
assumption of induction,
\begin{eqnarray}
&&(\sum_{\stackrel{{\{l_{*}\}\amalg\{m_{*}\}}=\{n_{*}\}/\{n_{l}\}}
{\{m_{*}\}\neq\emptyset}}
\sum_{d_{0}=0}^{d}\sum_{t_{1}=0}^{d-d_{0}}\cdots
\sum_{t_{ind(\{l_{*}\})+n_{l}-2}=0}^{t_{ind(\{l_{*}\})+n_{l}-3}}
C^{d-d_{0}}(\{l_{*}\}\cup\{n_{l}-1\}
;t_{1}, \cdots,t_{ind(\{l_{*}\})+n_{l}-2})\cdot\nonumber\\
&&\prod_{i=0}^{ind(\{l_{*}\})+n_{l}-2}
L_{n+1-a-i+(N-k+1)(d-d_{0}-t_{i})}^{N+1,k,t_{i}-t_{i+1}})
\cdot\nonumber\\
&&
(\sum_{u_{1}=0}^{d_{0}}\cdots
\sum^{u_{ind(\{l_{*}\})-1}}_{u_{ind(\{l_{*}\})}=0}
d_{0}C^{d_{0}}(\{l_{*}\}\cup\{n_{l}-1\}
;u_{1}, \cdots,u_{ind(\{m_{*}\})})\cdot\nonumber\\
&&\prod_{j=0}^{ind(\{m_{*}\})}
L_{n+1-a+1+ind(\{l_{*}\})-j+(N-k+1)(d-d_{0})
+(N-k+1)(d_{0}-u_{j})}^{N+1,k,u_{j}-u_{j+1}})
\nonumber\\
&&+\sum_{\stackrel{{\{l_{*}\}\amalg\{m_{*}\}}=\{n_{*}\}/\{n_{l}\}}
{\{m_{*}\}\neq\emptyset}}
\sum_{d_{0}=0}^{d}\sum_{t_{1}=0}^{d-d_{0}}\cdots
\sum^{t_{ind(\{n_{*}\})-2}}_{t_{ind(\{n_{*}\})-1}=0}
C^{d-d_{0}}(\{n_{*}\}/\{n_{l}\}\cup\{n_{l}-1\}
;t_{1}, \cdots,t_{ind(\{n_{*}\})-1})\cdot\nonumber\\
&&\prod_{i=0}^{ind(\{n_{*}\})-1}
L_{n+1-a-i+(N-k+1)(d-d_{0}-t_{i})}^{N+1,k,t_{i}-t_{i+1}}
\cdot L_{n+1-a-ind(\{n_{*}\})+(N-k+1)(d-d_{0})}^{N+1,k,t_{i}-t_{i+1}}
\end{eqnarray}
Then an appropriate change of $t_{i}$'s and $u_{i}$'s leads to
(\ref{facto}). Q.E.D.

This formula tells us that if we take $N-k$ fairly large, we can determine 
the form of recursive formula without subtle complexity. Our conjecture 
asserts that these formulas obtained works in $N-k\geq 2$ case for rational
curves of arbitrary degree and in $N-k=1$ case for curves whose degree 
is more than 2. 

Using  this, we  calculate some examples.
\begin{eqnarray}
&&C^{d}(\{2,2\};d_{1},d_{2})=d_{1}+d-d_{2}\nonumber\\ 
&&C^{d}(\{3,2\};d_{1},d_{2},d_{3})=d_{1}+d-d_{3}
\nonumber\\
&&C^{d}(\{3,3\};d_{1},d_{2},d_{3},d_{4})=d+d_{1}+d_{2}-d_{3}-d_{4}\nonumber\\
&&C^{d}(\{4,2\};d_{1},d_{2},d_{3},d_{4})=d+d_{1}-d_{4}\nonumber\\
&&C^{d}(\{2,2,2\};d_{1},d_{2},d_{3})=2d_{2}\cdot(d-d_{2})
+d_{1}\cdot(d_{1}+d_{2}-d_{3})
\nonumber\\
&&+(d-d_{3})\cdot(d+d_{1}-d_{2}-d_{3})\nonumber\\
&&C^{d}(\{3,2,2\};d_{1},d_{2},d_{3},d_{4})=d_{1}\cdot(d_{1}+d_{2}-d_{4})
+(d-d_{2})\cdot d_{2}
+(d-d_{3})\cdot d_{3}\nonumber\\
&&+(d-d_{4})\cdot(d+d_{1}-d_{3}-d_{4})\nonumber\\
&&C^{d}(\{2,2,2,2\};d_{1},d_{2},d_{3},d_{4})=d_{1}\cdot(2d_{3}\cdot
(d_{1}-d_{3})+d_{2}\cdot(d_{2}+d_{3}-d_{4})\nonumber\\
&&+(d_{1}-d_{4})\cdot(d_{1}+
d_{2}-d_{3}-d_{4}))
+3(d-d_{2})\cdot d_{2}\cdot(d_{2}+d_{3}-d_{4})\nonumber\\
&&+3(d-d_{3})\cdot d_{3}\cdot(d+d_{1}-d_{2}-d_{3})
+(d-d_{4})\cdot(2(d_{2}-d_{4})\cdot(d-d_{2})\nonumber\\
&&+(d_{1}-d_{4})\cdot(d_{1}+d_{2}-d_{3}-d_{4})   
+(d-d_{3})\cdot(d+d_{1}-d_{2}-d_{3}))\nonumber\\
\label{cre}
\end{eqnarray}
And specialization process can be systematically done by the
following formula.
\begin{eqnarray}
&&[A_{a_{1}-1}\cap H,A_{a_{2}-1}\cap H,\cdots,A_{a_{d+1}-1}\cap H;d,N+1,k]
\nonumber\\ 
&&=\sum_{m=1}^{d+1}\sum_{\stackrel{\amalg_{j=1}^{m}U_{j}=\{a_{*}\}}
{U_{j}\neq\emptyset}}(-1)^{d+1-m}
([A_{ind(U_{1})+1},A_{ind(U_{2})+1},\cdots,A_{ind(U_{m})+1};d,N+1,k]\cdot
\nonumber\\
&&\cdot(\prod_{j=1}^{m}({}^{\sharp}(U_{j})-1)!))
\label{sys}
\end{eqnarray}
Application of (\ref{sys}) leads us to,
\begin{eqnarray}
&&G[A_{a}^{N+1}\cap H, A_{b}^{N+1}\cap H, A_{1}^{N+1}\cap H,
A_{1}^{N+1}\cap H,A_{1}^{N+1}\cap H;4,N+1,k]
\nonumber\\
&=&[A_{a+1}^{N+1}, A_{b+1}^{N+1}, A_{2}^{N+1},A_{2}^{N+1}
A_{2}^{N+1};4,N+1,k]\nonumber\\
&-&3([A_{a+1}^{N+1}, A_{b+2}^{N+1}, A_{2}^{N+1}, A_{2}^{N+1};4,N+1,k]
\nonumber\\
&+&[A_{a+2}^{N+1}, A_{b+1}^{N+1}, A_{2}^{N+1}, A_{2}^{N+1};4,N+1,k])\nonumber\\
&-&3[A_{a+1}^{N+1}, A_{b+1}^{N+1}, A_{3}^{N+1}, A_{2}^{N+1};4,N+1,k]\nonumber\\
&+&6([A_{a+3}^{N+1}, A_{b+1}^{N+1}, A_{2}^{N+1};4,N+1,k]
+[A_{a+2}^{N+1}, A_{b+2}^{N+1}, A_{2}^{N+1};4,N+1,k]\nonumber\\
&+&[A_{a+1}^{N+1}, A_{b+3}^{N+1}, A_{2}^{N+1};4,N+1,k])\nonumber\\
&+&3([A_{a+2}^{N+1}, A_{b+1}^{N+1},A_{3}^{N+1};4,N+1,k]
+[A_{a+1}^{N+1}, A_{b+2}^{N+1},A_{3}^{N+1};4,N+1,k])\nonumber\\
&+&2[A_{a+1}^{N+1}, A_{b+1}^{N+1},A_{4}^{N+1};4,N+1,k])\nonumber\\
&-&6([A_{a+4}^{N+1}, A_{b+1}^{N+1};4,N+1,k]
+[A_{a+3}^{N+1}, A_{b+2}^{N+1};4,N+1,k]\nonumber\\
&+&[A_{a+2}^{N+1}, A_{b+3}^{N+1};4,N+1,k]
+[A_{a+1}^{N+1}, A_{b+4}^{N+1};4,N+1,k])\nonumber\\
&&(a+b=  N-3+4(N-k))\nonumber\\
&&\\
&&G[A_{a}^{N+1}\cap H, A_{b}^{N+1}\cap H, A_{1}^{N+1}\cap H,
A_{1}^{N+1}\cap H,A_{1}^{N+1}\cap H,
A_{1}^{N+1}\cap H;5,N+1,k]\nonumber\\
&=&[A_{a+1}^{N+1}, A_{b+1}^{N+1}, A_{2}^{N+1},A_{2}^{N+1}
A_{2}^{N+1},A_{2}^{N+1};5,N+1,k]\nonumber\\
&-&4([A_{a+1}^{N+1}, A_{b+2}^{N+1}, A_{2}^{N+1}, A_{2}^{N+1}, 
A_{2}^{N+1};5,N+1,k]\nonumber\\
&+&[A_{a+2}^{N+1}, A_{b+1}^{N+1}, A_{2}^{N+1}, A_{2}^{N+1},A_{2}^{N+1}
;5,N+1,k])\nonumber\\
&-&6[A_{a+1}^{N+1}, A_{b+1}^{N+1}, A_{3}^{N+1}, A_{2}^{N+1}, 
A_{2}^{N+1};5,N+1,k]\nonumber\\
&+&12([A_{a+3}^{N+1}, A_{b+1}^{N+1}, A_{2}^{N+1},A_{2}^{N+1};5,N+1,k]
\nonumber\\
&+&[A_{a+2}^{N+1}, A_{b+2}^{N+1}, A_{2}^{N+1},A_{2}^{N+1};5,N+1,k]
+[A_{a+1}^{N+1}, A_{b+3}^{N+1}, A_{2}^{N+1}, A_{2}^{N+1};5,N+1,k])\nonumber\\
&+&12([A_{a+2}^{N+1}, A_{b+1}^{N+1},A_{3}^{N+1}, A_{2}^{N+1};5,N+1,k]
\nonumber\\
&+&[A_{a+1}^{N+1}, A_{b+2}^{N+1},A_{3}^{N+1},A_{2}^{N+1};5,N+1,k])\nonumber\\
&+&8[A_{a+1}^{N+1}, A_{b+1}^{N+1},A_{4}^{N+1}, A_{2}^{N+1};5,N+1,k]\nonumber\\
&+&3[A_{a+1}^{N+1}, A_{b+1}^{N+1},A_{3}^{N+1}, A_{3}^{N+1};5,N+1,k]\nonumber\\
&-&24([A_{a+4}^{N+1}, A_{b+1}^{N+1},A_{2}^{N+1};5,N+1,k]
+[A_{a+3}^{N+1}, A_{b+2}^{N+1}, A_{2}^{N+1};5,N+1,k]\nonumber\\
&+&[A_{a+2}^{N+1}, A_{b+3}^{N+1}, A_{2}^{N+1};5,N+1,k]
+[A_{a+1}^{N+1}, A_{b+4}^{N+1},A_{2}^{N+1};5,N+1,k])\nonumber\\
&-&12([A_{a+3}^{N+1}, A_{b+1}^{N+1}, A_{3}^{N+1};5,N+1,k]
+[A_{a+2}^{N+1}, A_{b+2}^{N+1}, A_{3}^{N+1};5,N+1,k]\nonumber\\
&+&[A_{a+1}^{N+1}, A_{b+3}^{N+1}, A_{3}^{N+1};5,N+1,k])\nonumber\\
&-&8([A_{a+2}^{N+1}, A_{b+1}^{N+1},A_{4}^{N+1};5,N+1,k]
+[A_{a+1}^{N+1}, A_{b+2}^{N+1},A_{4}^{N+1};5,N+1,k])\nonumber\\
&-&6[A_{a+1}^{N+1}, A_{b+1}^{N+1},A_{5}^{N+1};5,N+1,k])\nonumber\\
&+&24([A_{a+5}^{N+1}, A_{b+1}^{N+1};5,N+1,k]
+[A_{a+4}^{N+1}, A_{b+2}^{N+1};5,N+1,k]\nonumber\\
&+&[A_{a+3}^{N+1}, A_{b+3}^{N+1};5,N+1,k]
+[A_{a+2}^{N+1}, A_{b+4}^{N+1};5,N+1,k]\nonumber\\
&+&[A_{a+1}^{N+1}, A_{b+4}^{N+1};5,N+1,k])\nonumber\\
&&(a+b=  N-3+5(N-k)).\nonumber\\
\label{sre}
\end{eqnarray}
By combining (\ref{sre}) with (\ref{cre}), we can obtain 
specialization results for quartics. 
\begin{eqnarray}
&&16L^{N,k,4}_{n}+R\nonumber\\
&=&\frac{3}{2}L_{n-3}^{4}+\frac{13}{2}L_{n-2}^{4}
+\frac{13}{2}L_{n-1}^{4}+\frac{3}{2}L_{n}^{4}\nonumber\\
&&+2L_{n-2}^{1}L_{n-2+N-k}^{3}+2L_{n-1}^{1}L_{n-2+N-k}^{3}
+6L_{n}^{1}L_{n-2+N-k}^{3}\nonumber\\
&&+8L_{n-1}^{1}L_{n-1+N-k}^{3}
+12L_{n}^{1}L_{n-1+N-k}^{3}+6L_{n}^{1}L_{n+N-k}^{3}\nonumber\\
&&+3L_{n-2}^{2}L_{n-1+2(N-k)}^{2}+7L_{n-1}^{2}L_{n-1+2(N-k)}^{2}
+6L_{n}^{2}L_{n-1+2(N-k)}^{2}\nonumber\\
&&+10L_{n-1}^{2}L_{n+2(N-k)}^{2}
+7L_{n}^{2}L_{n+2(N-k)}^{2}+3L_{n}^{2}L_{n+1+2(N-k)}^{2}\nonumber\\
&&+6L_{n-2}^{3}L_{n+3(N-k)}^{1}+12L_{n-1}^{3}L_{n+3(N-k)}^{1}
+6L_{n}^{3}L_{n+3(N-k)}^{1}\nonumber\\
&&+8L_{n-1}^{3}L_{n+1+3(N-k)}^{1}
+2L_{n}^{3}L_{n+1+3(N-k)}^{1}+2L_{n}^{3}L_{n+2+3(N-k)}^{1}\nonumber\\
&&+4L_{n-1}^{1}L_{n-1+N-k}^{1}L_{n-1+2(N-k)}^{2}
+9L_{n}^{1}L_{n-1+N-k}^{1}L_{n-1+2(N-k)}^{2}\nonumber\\
&&+10L_{n}^{1}L_{n+N-k}^{1}L_{n-1+2(N-k)}^{2}
+12L_{n}^{1}L_{n+N-k}^{1}L_{n+2(N-k)}^{2}\nonumber\\
&&+8L_{n-1}^{1}L_{n-1+N-k}^{2}L_{n+3(N-k)}^{1}
+14L_{n}^{1}L_{n-1+N-k}^{2}L_{n+3(N-k)}^{1}\nonumber\\
&&+14L_{n}^{1}L_{n+N-k}^{2}L_{n+3(N-k)}^{1}
+8L_{n}^{1}L_{n+N-k}^{2}L_{n+1+3(N-k)}^{1}\nonumber\\
&&+12L_{n-1}^{2}L_{n+2(N-k)}^{1}L_{n+3(N-k)}^{1}
+10L_{n}^{2}L_{n+2(N-k)}^{1}L_{n+3(N-k)}^{1}\nonumber\\
&&+9L_{n}^{2}L_{n+1+2(N-k)}^{1}L_{n+3(N-k)}^{1}
+4L_{n}^{2}L_{n+1+2(N-k)}^{1}L_{n+1+3(N-k)}^{1}\nonumber\\
&&+16L_{n}^{1}L_{n+N-k}^{1}L_{n+2(N-k)}^{1}L_{n+3(N-k)}^{1}
\label{trial}    
\end{eqnarray}
In (\ref{trial}), we omit $ N+1,k$ from $L_{n}^{N+1,k,d}$ 
for brevity. We will omit them from now on.
Next, we determine the contribution from connected reducible 
curves $R$ indirectly.  Assuming that this specialization result
exhausts all the terms that appear in ``true `` recursion relations 
(this is true for $d\leq 3$ rational curves), we  set unknown coefficients 
for terms of  
reducible curves considering symmetry of coefficients, which can be seen 
in the specialization results. 
\begin{eqnarray}
&&L^{N,k,4}_{n}\nonumber\\
&&=\frac{3}{32}L_{n-3}^{4}+\frac{13}{32}L_{n-2}^{4}
+\frac{13}{32}L_{n-1}^{4}+\frac{3}{32}L_{n}^{4}\nonumber\\
&&+a_{1}L_{n-2}^{1}L_{n-2+N-k}^{3}+a_{2}L_{n-1}^{1}L_{n-2+N-k}^{3}
+a_{3}L_{n}^{1}L_{n-2+N-k}^{3}\nonumber\\
&&+a_{4}L_{n-1}^{1}L_{n-1+N-k}^{3}
+a_{5}L_{n}^{1}L_{n-1+N-k}^{3}+a_{6}L_{n}^{1}L_{n+N-k}^{3}\nonumber\\
&&+b_{1}L_{n-2}^{2}L_{n-1+2(N-k)}^{2}+b_{2}L_{n-1}^{2}L_{n-1+2(N-k)}^{2}
+b_{3}L_{n}^{2}L_{n-1+2(N-k)}^{2}\nonumber\\
&&+b_{4}L_{n-1}^{2}L_{n+2(N-k)}^{2}
+b_{2}L_{n}^{2}L_{n+2(N-k)}^{2}+b_{1}L_{n}^{2}L_{n+1+2(N-k)}^{2}\nonumber\\
&&+a_{6}L_{n-2}^{3}L_{n+3(N-k)}^{1}+a_{5}L_{n-1}^{3}L_{n+3(N-k)}^{1}
+a_{3} L_{n}^{3}L_{n+3(N-k)}^{1}\nonumber\\
&&+a_{4}L_{n-1}^{3}L_{n+1+3(N-k)}^{1}
+a_{2}L_{n}^{3}L_{n+1+3(N-k)}^{1}+a_{1}L_{n}^{3}L_{n+2+3(N-k)}^{1}\nonumber\\
&&+c_{1}L_{n-1}^{1}L_{n-1+(N-k)}^{1}L_{n-1+2(N-k)}^{2}
+c_{2}L_{n}^{1}L_{n-1+N-k}^{1}L_{n-1+2(N-k)}^{2}\nonumber\\
&&+c_{3}L_{n}^{1}L_{n+N-k}^{1}L_{n-1+2(N-k)}^{2}
+c_{4}L_{n}^{1}L_{n+N-k}^{1}L_{n+2(N-k)}^{2}\nonumber\\
&&+d_{1}L_{n-1}^{1}L_{n-1+N-k}^{2}L_{n+3(N-k)}^{1}
+d_{2}L_{n}^{1}L_{n-1+N-k}^{2}L_{n+3(N-k)}^{1}\nonumber\\
&&+d_{2}L_{n}^{1}L_{n+N-k}^{2}L_{n+3(N-k)}^{1}
+d_{1}L_{n}^{1}L_{n+N-k}^{2}L_{n+1+3(N-k)}^{1}\nonumber\\
&&+c_{4}L_{n-1}^{2}L_{n+2(N-k)}^{1}L_{n+3(N-k)}^{1}
+c_{3}L_{n}^{2}L_{n+2(N-k)}^{1}L_{n+3(N-k)}^{1}\nonumber\\
&&+c_{2}L_{n}^{2}L_{n+1+2(N-k)}^{1}L_{n+3(N-k)}^{1}
+c_{1}L_{n}^{2}L_{n+1+2(N-k)}^{1}L_{n+1+3(N-k)}^{1}\nonumber\\
&&+e_{1}L_{n}^{1}L_{n+N-k}^{1}L_{n+2(N-k)}^{1}L_{n+3(N-k)}^{1}    
\end{eqnarray}
From conjectural characteristics of recursion relation 
that imposes $\gamma_{0}^{N,k,4}=\gamma_{0}^{N+1,k,4}$, we 
obtain some constraints on these unknown coefficients. 
\begin{eqnarray}
&&a_{3}=\frac{2}{9},\;\;a_{2}+a_{5}
=\frac{7}{9},\;\; a_{1}+a_{4}+a_{6}=1\nonumber\\
&&b_{3}=\frac{1}{4},\;\;2b_{2}=\frac{3}{4},\;\;2b_{1}+b_{4}=1\nonumber\\
&&c_{2}=\frac{1}{3},\;\;c_{3}=\frac{1}{2},\;\;c_{1}+c_{4}=1\nonumber\\
&&2d_{1}=1,\;\;d_{2}=d_{3}=\frac{2}{3}\nonumber\\
&&e_{1}=1
\label{const}
\end{eqnarray}
If we compare (\ref{trial}) with (\ref{const}), we can see $ a_{1}+a_{4}
+a_{6}=1,2b_{1}+b_{4}=1,,c_{1}+c_{4}=1,2d_{1}=1, e_{1}=1$ are 
automatically satisfied in (\ref{trial}). And we heuristically set 
$a_{1}=\frac{1}{8},a_{4}=\frac{1}{2},a_{6}=\frac{3}{8}, 
b_{1}=\frac{3}{16},b_{4}=\frac{5}{8},c_{1}=\frac{1}{4},
c_{4}=\frac{3}{4},d_{1}=\frac{1}{2},
e_{1}=1$.
 We have to 
note some combinatorial relation on these coefficients that 
can be seen from (\ref{trial}).
\begin{eqnarray}
&&\frac{3}{32}x^{3}+\frac{13}{32}x^{2}y+\frac{13}{32}xy^{2}
+\frac{3}{32}y^{3}=(\frac{3x+y}{4})(\frac{2x+2y}{4})(\frac{x+3y}{4})
\nonumber\\
&&\frac{1}{8}x^{2}+\frac{1}{2}xy+\frac{3}{8}y^{2}=(\frac{2x+2y}{4})
(\frac{x+3y}{4})
,\;\;\frac{3}{16}x^{2}+\frac{5}{8}xy+\frac{3}{16}y^{2}=(\frac{3x+y}{4})
(\frac{x+3y}{4})\nonumber\\
&&\frac{3}{8}x^{2}+\frac{1}{2}xy+\frac{1}{8}y^{2}=(\frac{3x+y}{4})
(\frac{2x+2y}{4})\nonumber\\
&&\frac{1}{4}x+\frac{3}{4}y=(\frac{x+3y}{4}),\;\;\;
\frac{1}{2}x+\frac{1}{2}y=(\frac{2x+2y}{4})\nonumber\\
&&\frac{3}{4}x+\frac{1}{4}y=(\frac{3x+y}{4})\nonumber\\
\end{eqnarray}
As a summary of discussion given so far, we propose the following.
\begin{conj}
The prototype result obtained from specialization 
exhausts all the addends that appear in the ``true''  
recursive formula and coefficients described by the following 
generating polynomial remain unchanged after subtraction of 
contribution from ``$R$'' term.
\end{conj}
\begin{equation}
\prod_{j=1}^{d-1}(\frac{jx+(d-j)y}{d}+z_{j})
\end{equation}
{\bf Examples}
\begin{eqnarray}
d=2\;\;&&(\frac{x+y}{2})+z_{1}\nonumber\\
d=3\;\;&&(\frac{2x^{2}+5xy+2y^{2}}{9})
          +(\frac{2x+y}{3})z_{1}
          +(\frac{x+2y}{3})z_{2}
          +z_{1}z_{2}\nonumber\\
d=4\;\;&&(\frac{3x^{3}+13x^{2}y+13xy^{2}+3y^{3}}{32})\nonumber\\
       &&+(\frac{x^{2}+4xy+3y^{2}}{8})z_{3}
         +(\frac{3x^{2}+10xy+3y^{2}}{16})z_{2}
         +(\frac{3x^{2}+4xy+y^{2}}{8})z_{1}\nonumber\\
       &&+(\frac{3x+y}{4})z_{1}z_{2}
         +(\frac{x+y}{2})z_{1}z_{3}
         +(\frac{x+3y}{4})z_{2}z_{3}\nonumber\\
       &&+z_{1}z_{2}z_{3}\nonumber\\
d=5\;\;&&(\frac{24x^{4}+154x^{3}y+269x^{2}y^{2}+154xy^{3}+24y^{4}}{625})
\nonumber\\
&&+(\frac{6x^{3}+37x^{2}y+58xy^{2}+24y^{3}}{125})z_{4}
+(\frac{8x^{3}+46x^{2}y+59xy^{2}+12y^{3}}{125})z_{3}\nonumber\\
&&+(\frac{12x^{3}+59x^{2}y+46xy^{2}+8y^{3}}{125})z_{2}
+(\frac{24x^{3}+58x^{2}y+37xy^{2}+6y^{3}}{125})z_{1}\nonumber\\
&&+(\frac{2x^{2}+11xy+12y^{2}}{25})z_{3}z_{4}
+(\frac{6x^{2}+13xy+6y^{2}}{25})z_{1}z_{4}
+(\frac{12x^{2}+11xy+2y^{2}}{25})z_{1}z_{2}\nonumber\\
&&+(\frac{3x^{2}+14xy+8y^{2}}{25})z_{2}z_{4}
+(\frac{4x^{2}+17xy+4y^{2}}{25})z_{2}z_{3}
+(\frac{8x^{2}+14xy+3y^{2}}{25})z_{1}z_{3}\nonumber\\
&&+(\frac{4x+y}{5})z_{1}z_{2}z_{3}
  +(\frac{3x+2y}{5})z_{1}z_{2}z_{4}
  +(\frac{2x+3y}{5})z_{1}z_{3}z_{4}
  +(\frac{x+4y}{5})z_{2}z_{3}z_{4}\nonumber\\
&&+z_{1}z_{2}z_{3}z_{4}\nonumber\\
\label{poly}
\end{eqnarray}
Then we go back to the argument of quartic curves.
 Remaining unknown coefficient is only $a_{2}$ ($a_{5}$). 
Then we use some numerical results obtained from torus action method. 
\begin{eqnarray}
H_{q,e}^{*}(M_{9}^{7})&&\nonumber\\
{\cal O}_{e}\cdot{\cal O}_{e}&=&{\cal O}_{e^{2}}+5040q\nonumber\\
{\cal O}_{e}\cdot{\cal O}_{e^{2}}&=&{\cal O}_{e^{3}}
+56196{\cal O}_{e}q\nonumber\\
{\cal O}_{e}\cdot{\cal O}_{e^{3}}&=&{\cal O}_{e^{4}}
+200452{\cal O}_{e^{2}}q+2056259520q^{2}\nonumber\\
{\cal O}_{e}\cdot{\cal O}_{e^{4}}&=& {\cal O}_{e^{5}}
+300167{\cal O}_{e^{3}}q
+24699506832{\cal O}_{e}q^{2}\nonumber\\
{\cal O}_{e}\cdot{\cal O}_{e^{5}}&=&
{\cal O}_{e^{6}}
+200452{\cal O}_{e^{4}}q
+53751685624{\cal O}_{e^{2}}q^{2}\nonumber\\
&&+534155202302400q^{3}\nonumber\\
{\cal O}_{e}\cdot{\cal O}_{e^{6}}&=&{\cal O}_{e^{7}}
+56196{\cal O}_{e^{5}}q
+24699506832{\cal O}_{e^{3}}q^{2}\nonumber\\
&&+1920365635990032{\cal O}_{e}q^{3}\nonumber\\
{\cal O}_{e}\cdot{\cal O}_{e^{7}}&=&
5040{\cal O}_{e^{6}}q
+2056259520{\cal O}_{e^{4}}q^{2}\nonumber\\
&&+534155202302400{\cal O}_{e^{2}}q^{3}\nonumber\\
&&+5112982794486067200q^{4}\nonumber\\
\end{eqnarray}
\begin{eqnarray}
H_{q,e}^{*}(M_{8}^{7})&&\nonumber\\
{\cal O}_{e}\cdot{\cal O}_{e}&=&{\cal O}_{e^{2}}+51156{\cal O}_{e}q
+ 1311357600 q^{2}\nonumber\\
{\cal O}_{e}\cdot{\cal O}_{e^{2}}&=&{\cal O}_{e^{3}}
+ 195412{\cal O}_{e^{2}}q+24642483768{\cal O}_{e}q^{2} 
+ 675477943761600q^{3}\nonumber\\
{\cal O}_{e}\cdot{\cal O}_{e^{3}}&=&{\cal O}_{e^{4}}
+ 295127{\cal O}_{e^{3}}q+99394671712{\cal O}_{e^{2}}q^{2}
+12622841688846312{\cal O}_{e}q^{3}\nonumber\\
&&+352826466584918860800q^{4}\nonumber\\
{\cal O}_{e}\cdot{\cal O}_{e^{4}}&=& {\cal O}_{e^{5}}
+195412{\cal O}_{e^{4}}q
+99394671712{\cal O}_{e^{3}}q^{2}
+32755090390395744{\cal O}_{e^{2}}q^{3}\nonumber\\
&&+4092145211387781662688{\cal O}_{e}q^{4}+\cdots\nonumber\\
{\cal O}_{e}\cdot{\cal O}_{e^{5}}&=&
{\cal O}_{e^{6}}
+51156{\cal O}_{e^{5}}q
+24642483768{\cal O}_{e^{4}}q^{2}
+12622841688846312{\cal O}_{e^{3}}q^{3}\nonumber\\
&&+4092145211387781662688{\cal O}_{e^{2}}q^{4}+\cdots \nonumber\\
{\cal O}_{e}\cdot{\cal O}_{e^{6}}&=&
1311357600{\cal O}_{e^{5}}q^{2}
+675477943761600{\cal O}_{e^{4}}q^{3}\nonumber\\
&&+352826466584918860800{\cal O}_{e^{3}}q^{4}+\cdots\nonumber\\
\end{eqnarray}
Then applying the recursion relation with unknown $a_{2}$ 
for 352826466584918860800, we 
obtain the following equation. 
\begin{equation}
\frac{109466}{3}= (\frac{7}{9}-a_{2})\cdot5040+a_{2}\cdot200452
\end{equation}
And we have $a_{2}=\frac{1}{6}$.
The final result is,     
\begin{eqnarray}
L_{n}^{N,k,4}
&=&\frac{1}{32}(3L_{n-3}^{4}+13L_{n-2}^{4}
+13L_{n-1}^{4}+3L_{n}^{4})\nonumber\\
&&+\frac{1}{72}(9L_{n-2}^{1}L_{n-2+N-k}^{3}+12L_{n-1}^{1}L_{n-2+N-k}^{3}
+16L_{n}^{1}L_{n-2+N-k}^{3}\nonumber\\
&&+36L_{n-1}^{1}L_{n-1+N-k}^{3}
+44L_{n}^{1}L_{n-1+N-k}^{3}+27L_{n}^{1}L_{n+N-k}^{3})\nonumber\\
&&+\frac{1}{16}(3L_{n-2}^{2}L_{n-1+2(N-k)}^{2}
+6L_{n-1}^{2}L_{n-1+2(N-k)}^{2}
+4L_{n}^{2}L_{n-1+2(N-k)}^{2}\nonumber\\
&&+10L_{n-1}^{2}L_{n+2(N-k)}^{2}
+6L_{n}^{2}L_{n+2(N-k)}^{2}+3L_{n}^{2}L_{n+1+2(N-k)}^{2})
\nonumber\\
&&+\frac{1}{72}(27L_{n-2}^{3}L_{n+3(N-k)}^{1}
+44L_{n-1}^{3}L_{n+3(N-k)}^{1}
+16L_{n}^{3}L_{n+3(N-k)}^{1}\nonumber\\
&&+36L_{n-1}^{3}L_{n+1+3(N-k)}^{1}
+12L_{n}^{3}L_{n+1+3(N-k)}^{1}+
9L_{n}^{3}L_{n+2+3(N-k)}^{1})\nonumber\\
&&+\frac{1}{12}(3L_{n-1}^{1}L_{n-1+N-k}^{1}L_{n-1+2(N-k)}^{2}
+4L_{n}^{1}L_{n-1+N-k}^{1}L_{n-1+2(N-k)}^{2}\nonumber\\
&&+6L_{n}^{1}L_{n+N-k}^{1}L_{n-1+2(N-k)}^{2}
+9L_{n}^{1}L_{n+N-k}^{1}L_{n+2(N-k)}^{2})\nonumber\\
&&+\frac{1}{6}(3L_{n-1}^{1}L_{n-1+N-k}^{2}L_{n+3(N-k)}^{1}
+4L_{n}^{1}L_{n-1+N-k}^{2}L_{n+3(N-k)}^{1}\nonumber\\
&&+4L_{n}^{1}L_{n+N-k}^{2}L_{n+3(N-k)}^{1}
+3L_{n}^{1}L_{n+N-k}^{2}L_{n+1+3(N-k)}^{1})\nonumber\\
&&+\frac{1}{12}(9L_{n-1}^{2}L_{n+2(N-k)}^{1}L_{n+3(N-k)}^{1}
+6L_{n}^{2}L_{n+2(N-k)}^{1}L_{n+3(N-k)}^{1}\nonumber\\
&&+4L_{n}^{2}L_{n+1+2(N-k)}^{1}L_{n+3(N-k)}^{1}
+3L_{n}^{2}L_{n+1+2(N-k)}^{1}L_{n+1+3(N-k)}^{1})\nonumber\\
&&+L_{n}^{1}L_{n+N-k}^{1}L_{n+2(N-k)}^{1}L_{n+3(N-k)}^{1}.    
\end{eqnarray}
Of course, we have the following equality.
\begin{eqnarray} 
&& (13/32)\cdot5112982794486067200\nonumber\\
&+&(1/6)\cdot5040\cdot534155202302400
+(2/9)\cdot56196\cdot534155202302400\nonumber\\
&+&(1/2)\cdot5040\cdot1920365635990032
+(11/18)\cdot56196\cdot1920365635990032\nonumber\\
&+&(3/8)\cdot56196\cdot534155202302400
+(3/8)\cdot2056259520\cdot53751685624\nonumber\\
&+&(1/4)\cdot24699506832\cdot53751685624
+(5/8)\cdot2056259520\cdot24699506832\nonumber\\
&+&(3/8)\cdot24699506832\cdot24699506832
+(3/16)\cdot24699506832\cdot2056259520\nonumber\\
&+&(11/18)\cdot534155202302400\cdot200452
+(2/9)\cdot1920365635990032\cdot200452\nonumber\\
&+&(1/2)\cdot534155202302400\cdot56196
+(1/6)\cdot1920365635990032\cdot56196\nonumber\\
&+&(1/8)\cdot1920365635990032\cdot5040
+(1/4)\cdot5040\cdot56196\cdot53751685624\nonumber\\
&+&(1/3)\cdot56196\cdot56196\cdot53751685624
+(1/2)\cdot56196\cdot200452\cdot53751685624\nonumber\\
&+&(3/4)\cdot56196\cdot200452\cdot24699506832
+(1/2)\cdot5040\cdot24699506832\cdot200452\nonumber\\
&+&(2/3)\cdot56196\cdot24699506832\cdot200452
+(2/3)\cdot56196\cdot53751685624\cdot200452\nonumber\\
&+&(1/2)\cdot56196\cdot53751685624\cdot56196
+(3/4)\cdot2056259520\cdot300167\cdot200452\nonumber\\
&+&(1/2)\cdot24699506832\cdot300167\cdot200452
+(1/3)\cdot24699506832\cdot200452\cdot200452\nonumber\\
&+&(1/4)\cdot24699506832\cdot200452\cdot56196
+ 56196\cdot200452\cdot300167\cdot200452\nonumber\\
&=& 4092145211387781662688
\end{eqnarray}
We further checked numerically the previous conjecture of 
virtual quantum cohomology ring of Calabi-Yau hypersurfaces in $CP^{k-1}$.
\begin{eqnarray}
k^{4k}a_{4}&=&\tilde{L}_{0}^{k,k,4}\nonumber\\
k^{4k}b_{4}&=&\frac{1}{4}\tilde{L}_{1}^{k,k,4}+               
\frac{1}{3}\tilde{L}_{1}^{k,k,3}\cdot\tilde{L}_{0}^{k,k,1}+
\frac{1}{2}\tilde{L}_{1}^{k,k,2}\cdot\tilde{L}_{0}^{k,k,2}
+\tilde{L}_{1}^{k,k,1}\cdot\tilde{L}_{0}^{k,k,3}.       
\end{eqnarray}
Moreover, we obtained correct Gromov-Witten invariant 
for $M_{5}^{5}$ case under the assumption of (\ref{tra}).
To write out general recursive formula is tedious, and we give the result for 
$M_{5}^{5}$ in the following. 
\begin{eqnarray}
&&(c_4-b_4)-c_3\cdot b_1+2\cdot b_1\cdot b_3-b_3\cdot c_1-c_2\cdot b_2
\nonumber\\
&&+c_2\cdot(b_1)^{2}-
3\cdot b_2\cdot (b_1)^{2}+2\cdot b_2\cdot b_1\cdot c_1
-c_1\cdot (b_1)^{3}+(b_1)^{4}+(b_2)^{2}\nonumber\\
&&+(c_1-b_1)\cdot(-(1/3)\cdot b_3+(3/2)\cdot b_1\cdot b_2-
(3/2)\cdot (b_1)^{3})\nonumber\\
&&+2\cdot(c_2-b_2-c_1\cdot b_1+(b_1)^{2})\cdot(-(1/2)\cdot b_2+(b_1)^{2})
\nonumber\\
&&+3\cdot(c_3-b_3-c_2\cdot b_1+2\cdot b_2\cdot b_1
-c_1\cdot b_2+c_1\cdot(b_1)^{2}-(b_1)^{3})\cdot(-b_1)
\nonumber\\
&&+(c_1-b_1)\cdot(-b_1)\cdot(-(1/2)\cdot b_2+(b_1)^{2})+
2\cdot(c_2-b_2-c_1\cdot b_1+(b_1)^{2})\cdot((b_1)^{2})
\nonumber\\
&&-(1/6)\cdot(c_1-b_1)\cdot((b_1)^{3})
\nonumber\\
&=& 3103585359375
\nonumber\\ 
&&\mbox{where}
\nonumber\\
&&b_1=770,\; b_2=1435650,\; b_3=3225308000,\; b_4=7894629141250\nonumber\\
&&c_1=1345,\; c_2=3296525,\; c_3=8940963625,\; c_4=25306794813125\nonumber
\end{eqnarray}

{\bf Remark 1}

Instead of using numerical results, we can derive $a_{2}= \frac{1}{6}$
from information of coefficients of hypergeometric series. We will 
explain it in determination of recursive formula for quintic curves.

{\bf Remark 2}

 We can formally generalize the polynomial description 
of recursive formulas (\ref{poly}) to include all the terms 
that appear in recursive formulas. For example, 
\begin{eqnarray}
d=3\;\;&&(\frac{2x^{2}+5xy+2y^{2}}{9})
          +(\frac{2x+y}{3}+\frac{1}{2}z_{1})z_{1}
          +(\frac{x+2y}{3}+\frac{1}{2}z_{2})z_{2}
          +z_{1}z_{2}\nonumber\\
d=4\;\;&&(\frac{3x^{3}+13x^{2}y+13xy^{2}+3y^{3}}{32})\nonumber\\
       &&+(\frac{x^{2}+4xy+3y^{2}}{8}
          +(\frac{3x+11y}{18})z_{3}
           +\frac{2}{9}(z_{3})^{2})z_{3}\nonumber\\
       &&+(\frac{3x^{2}+10xy+3y^{2}}{16}+(\frac{3x+3y}{8})z_{2} 
         +\frac{1}{4}(z_{2})^{2})z_{2}\nonumber\\
       &&+(\frac{3x^{2}+4xy+y^{2}}{8}+(\frac{11x+3y}{18})z_{1}
           +\frac{2}{9}(z_{1})^{2})z_{1}\nonumber\\
       &&+(\frac{3x+y}{4}+\frac{1}{2}z_{1}+\frac{1}{3}z_{2})z_{1}z_{2}
         +(\frac{x+y}{2}+\frac{2}{3}z_{1}+\frac{2}{3}z_{3})
         z_{1}z_{3}\nonumber\\
       &&+(\frac{x+3y}{4}+\frac{1}{3}z_{2}+\frac{1}{2}z_{3})
         z_{2}z_{3}\nonumber\\
       &&+z_{1}z_{2}z_{3}
\label{poly2}
\end{eqnarray}
From this formula, we can speculate that the number of 
addends in recursive formula for degree $d$ curves is equal to 
the number of degree $d-1$ monomials of $d+1$ variables, i.e., 
${}_{2d-1}C_{d}$. But this generalization does not have 
simple factorization property like (\ref{poly}). $\Box$

Then  we go on to determination of recursive formula for 
 quintic curves. 
The prototype result 
from specialization approach is the following.
\begin{eqnarray}
&&125L^{N,k,5}_{n}+R \nonumber\\
&=&\frac{24}{5}L_{n-4}^{5}+\frac{154}{5}L_{n-3}^{5}
+\frac{269}{5}L_{n-2}^{5}+\frac{154}{5}L_{n-1}^{5}
+\frac{24}{5}L_{n}^{5} \nonumber\\
&&+6L_{n-3}^{1}L_{n-3+N-k}^{4}+9L_{n-2}^{1}L_{n-3+N-k}^{4}
+2L_{n-1}^{1}L_{n-3+N-k}^{4}\nonumber\\
&&+24L_{n}^{1}L_{n-3+N-k}^{4}
+37L_{n-2}^{1}L_{n-2+N-k}^{4}
+38L_{n-1}^{1}L_{n-2+N-k}^{4}\nonumber\\
&&+80L_{n}^{1}L_{n-2+N-k}^{4}+58L_{n-1}^{1}L_{n-1+N-k}^{4}
+72L_{n}^{1}L_{n-1+N-k}^{4}\nonumber\\
&&+24L_{n}^{1}L_{n+N-k}^{4}\nonumber\\
&&+8L_{n-3}^{2}L_{n-2+2(N-k)}^{3}+12L_{n-2}^{2}L_{n-2+2(N-k)}^{3}
+28L_{n-1}^{2}L_{n-2+2(N-k)}^{3}\nonumber\\
&&+24L_{n}^{2}L_{n-2+2(N-k)}^{3} 
+46L_{n-2}^{2}L_{n-1+2(N-k)}^{3}
+74L_{n-1}^{2}L_{n-1+2(N-k)}^{3}\nonumber\\
&&+54L_{n}^{2}L_{n-1+2(N-k)}^{3}+59L_{n-1}^{2}L_{n+2(N-k)}^{3}
+33L_{n}^{2}L_{n+2(N-k)}^{3}\nonumber\\
&&+12L_{n}^{2}L_{n+1+2(N-k)}^{3}\nonumber\\
&&+12L_{n-3}^{3}L_{n-1+3(N-k)}^{2}+33L_{n-2}^{3}L_{n-1+3(N-k)}^{2}
+54L_{n-1}^{3}L_{n-1+3(N-k)}^{2}\nonumber\\
&&+24L_{n}^{3}L_{n-1+3(N-k)}^{2}
+59L_{n-2}^{3}L_{n+3(N-k)}^{2}
+74L_{n-1}^{3}L_{n+3(N-k)}^{2}\nonumber\\
&&+28L_{n}^{3}L_{n+3(N-k)}^{2}+46L_{n-1}^{3}L_{n+1+3(N-k)}^{2}
+12L_{n}^{3}L_{n+1+3(N-k)}^{2}\nonumber\\
&&+8L_{n}^{3}L_{n+2+3(N-k)}^{2}\nonumber\\
&&+24L_{n-3}^{4}L_{n+4(N-k)}^{1}+72L_{n-2}^{4}L_{n+4(N-k)}^{1}
+80L_{n-1}^{4}L_{n+4(N-k)}^{1}\nonumber\\
&&+24L_{n}^{4}L_{n+4(N-k)}^{1} 
+58L_{n-2}^{4}L_{n+1+4(N-k)}^{1}
+38L_{n-1}^{4}L_{n+1+4(N-k)}^{1}\nonumber\\
&&+2L_{n}^{4}L_{n+1+4(N-k)}^{1}+37L_{n-1}^{4}L_{n+2+4(N-k)}^{1}
+9L_{n}^{4}L_{n+2+4(N-k)}^{1}\nonumber\\
&&+6L_{n}^{4}L_{n+3+4(N-k)}^{1}\nonumber\\
&&+10L_{n-2}^{1}L_{n-2+N-k}^{1}L_{n-2+2(N-k)}^{3}
+3L_{n-1}^{1}L_{n-2+N-k}^{1}L_{n-2+2(N-k)}^{3}\nonumber\\
&&+32L_{n}^{1}L_{n-2+N-k}^{1}L_{n-2+2(N-k)}^{3}
+14L_{n-1}^{1}L_{n-1+N-k}^{1}L_{n-2+2(N-k)}^{3}\nonumber\\
&&+43L_{n}^{1}L_{n-1+N-k}^{1}L_{n-2+2(N-k)}^{3}
+42L_{n}^{1}L_{n+N-k}^{1}L_{n-2+2(N-k)}^{3}\nonumber\\
&&+55L_{n-1}^{1}L_{n-1+N-k}^{1}L_{n-1+2(N-k)}^{3}
+96L_{n}^{1}L_{n-1+N-k}^{1}L_{n-1+2(N-k)}^{3}\nonumber\\
&&+93L_{n}^{1}L_{n+N-k}^{1}L_{n-1+2(N-k)}^{3}
+60L_{n}^{1}L_{n+N-k}^{1}L_{n+2(N-k)}^{3}\nonumber\\
&&+30L_{n-2}^{1}L_{n-2+N-k}^{3}L_{n+4(N-k)}^{1}
+29L_{n-1}^{1}L_{n-2+N-k}^{3}L_{n+4(N-k)}^{1}\nonumber\\
&&+78L_{n}^{1}L_{n-2+N-k}^{3}L_{n+4(N-k)}^{1}
+86L_{n-1}^{1}L_{n-1+N-k}^{3}L_{n+4(N-k)}^{1}\nonumber\\
&&+123L_{n}^{1}L_{n-1+N-k}^{3}L_{n+4(N-k)}^{1}
+78L_{n}^{1}L_{n+N-k}^{3}L_{n+4(N-k)}^{1}\nonumber\\
&&+65L_{n-1}^{1}L_{n-1+N-k}^{3}L_{n+1+4(N-k)}^{1}
+86L_{n}^{1}L_{n-1+N-k}^{3}L_{n+1+4(N-k)}^{1}\nonumber\\
&&+29L_{n}^{1}L_{n+N-k}^{3}L_{n+1+4(N-k)}^{1}
+30L_{n}^{1}L_{n+N-k}^{3}L_{n+2+4(N-k)}^{1}\nonumber\\
&&+60L_{n-2}^{3}L_{n+3(N-k)}^{1}L_{n+4(N-k)}^{1}
+93L_{n-1}^{3}L_{n+3(N-k)}^{1}L_{n+4(N-k)}^{1}\nonumber\\
&&+42L_{n}^{3}L_{n+3(N-k)}^{1}L_{n+4(N-k)}^{1}
+96L_{n-1}^{3}L_{n+1+3(N-k)}^{1}L_{n+4(N-k)}^{1}\nonumber\\
&&+43L_{n}^{3}L_{n+1+3(N-k)}^{1}L_{n+4(N-k)}^{1}
+32L_{n}^{3}L_{n+2+3(N-k)}^{1}L_{n+4(N-k)}^{1}\nonumber\\
&&+55L_{n-1}^{3}L_{n+1+3(N-k)}^{1}L_{n+1+4(N-k)}^{1}
+14L_{n}^{3}L_{n+1+3(N-k)}^{1}L_{n+1+4(N-k)}^{1}\nonumber\\
&&+3L_{n}^{3}L_{n+2+3(N-k)}^{1}L_{n+1+4(N-k)}^{1}
+10L_{n}^{3}L_{n+2+3(N-k)}^{1}L_{n+2+4(N-k)}^{1}\nonumber\\
&&+15L_{n-2}^{1}L_{n-2+N-k}^{2}L_{n-1+3(N-k)}^{2}
+10L_{n-1}^{1}L_{n-2+N-k}^{2}L_{n-1+3(N-k)}^{2}\nonumber\\
&&+48L_{n}^{1}L_{n-2+N-k}^{2}L_{n-1+3(N-k)}^{2}
+42L_{n-1}^{1}L_{n-1+N-k}^{2}L_{n-1+3(N-k)}^{2}\nonumber\\
&&+76L_{n}^{1}L_{n-1+N-k}^{2}L_{n-1+3(N-k)}^{2}
+60L_{n}^{1}L_{n+N-k}^{2}L_{n-1+3(N-k)}^{2}\nonumber\\
&&+70L_{n-1}^{1}L_{n-1+N-k}^{2}L_{n+3(N-k)}^{2}
+104L_{n}^{1}L_{n-1+N-k}^{2}L_{n+3(N-k)}^{2}\nonumber\\
&&+76L_{n}^{1}L_{n+N-k}^{2}L_{n+3(N-k)}^{2}
+40L_{n}^{1}L_{n+N-k}^{2}L_{n+1+3(N-k)}^{2}\nonumber\\
&&+20L_{n-2}^{2}L_{n-1+2(N-k)}^{1}L_{n-1+3(N-k)}^{2}
+44L_{n-1}^{2}L_{n-1+2(N-k)}^{1}L_{n-1+3(N-k)}^{2}\nonumber\\
&&+36L_{n}^{2}L_{n-1+2(N-k)}^{1}L_{n-1+3(N-k)}^{2}
+59L_{n-1}^{2}L_{n+2(N-k)}^{1}L_{n-1+3(N-k)}^{2}\nonumber\\
&&+45L_{n}^{2}L_{n+2(N-k)}^{1}L_{n-1+3(N-k)}^{2}
+36L_{n}^{2}L_{n+1+2(N-k)}^{1}L_{n-1+3(N-k)}^{2}\nonumber\\
&&+85L_{n-1}^{2}L_{n+2(N-k)}^{1}L_{n+3(N-k)}^{2}
+59L_{n}^{2}L_{n+2(N-k)}^{1}L_{n+3(N-k)}^{2}\nonumber\\
&&+44L_{n}^{2}L_{n+1+2(N-k)}^{1}L_{n+3(N-k)}^{2}
+20L_{n}^{2}L_{n+1+2(N-k)}^{1}L_{n+1+3(N-k)}^{2}\nonumber\\
&&+40L_{n-2}^{2}L_{n-1+2(N-k)}^{2}L_{n+4(N-k)}^{1}
+76L_{n-1}^{2}L_{n-1+2(N-k)}^{2}L_{n+4(N-k)}^{1}\nonumber\\
&&+60L_{n}^{2}L_{n-1+2(N-k)}^{2}L_{n+4(N-k)}^{1}
+104L_{n-1}^{2}L_{n+2(N-k)}^{2}L_{n+4(N-k)}^{1}\nonumber\\
&&+76L_{n}^{2}L_{n+2(N-k)}^{2}L_{n+4(N-k)}^{1}
+48L_{n}^{2}L_{n+1+2(N-k)}^{2}L_{n+4(N-k)}^{1}\nonumber\\
&&+70L_{n-1}^{2}L_{n+2(N-k)}^{2}L_{n+1+4(N-k)}^{1}
+42L_{n}^{2}L_{n+2(N-k)}^{2}L_{n+1+4(N-k)}^{1}\nonumber\\
&&+10L_{n}^{2}L_{n+1+2(N-k)}^{2}L_{n+1+4(N-k)}^{1}
+15L_{n}^{2}L_{n+1+2(N-k)}^{2}L_{n+2+4(N-k)}^{1}\nonumber\\
&&+25L_{n-1}^{1}L_{n-1+N-k}^{1}L_{n-1+2(N-k)}^{1}L_{n-1+3(N-k)}^{2}
+64L_{n}^{1}L_{n-1+N-k}^{1}L_{n-1+2(N-k)}^{1}L_{n-1+3(N-k)}^{2}\nonumber\\
&&+63L_{n}^{1}L_{n+N-k}^{1}L_{n-1+2(N-k)}^{1}L_{n-1+3(N-k)}^{2}
+76L_{n}^{1}L_{n+N-k}^{1}L_{n+2(N-k)}^{1}L_{n-1+3(N-k)}^{2}\nonumber\\
&&+100L_{n}^{1}L_{n+N-k}^{1}L_{n+2(N-k)}^{1}L_{n+3(N-k)}^{2}
+50L_{n-1}^{1}L_{n-1+N-k}^{1}L_{n-1+2(N-k)}^{2}L_{n+4(N-k)}^{1}\nonumber\\
&&+101L_{n}^{1}L_{n-1+N-k}^{1}L_{n-1+2(N-k)}^{2}L_{n+4(N-k)}^{1}
+102L_{n}^{1}L_{n+N-k}^{1}L_{n-1+2(N-k)}^{2}L_{n+4(N-k)}^{1}\nonumber\\
&&+122L_{n}^{1}L_{n+N-k}^{1}L_{n+2(N-k)}^{2}L_{n+4(N-k)}^{1}
+75L_{n}^{1}L_{n+N-k}^{1}L_{n+2(N-k)}^{2}L_{n+1+4(N-k)}^{1}\nonumber\\
&&+75L_{n-1}^{1}L_{n-1+N-k}^{2}L_{n+3(N-k)}^{1}L_{n+4(N-k)}^{1}
+122L_{n}^{1}L_{n-1+N-k}^{2}L_{n+3(N-k)}^{1}L_{n+4(N-k)}^{1}\nonumber\\
&&+102L_{n}^{1}L_{n+N-k}^{2}L_{n+3(N-k)}^{1}L_{n+4(N-k)}^{1}
+101L_{n}^{1}L_{n+N-k}^{2}L_{n+1+3(N-k)}^{1}L_{n+4(N-k)}^{1}\nonumber\\
&&+50L_{n}^{1}L_{n+N-k}^{2}L_{n+1+3(N-k)}^{1}L_{n+1+4(N-k)}^{1}
+100L_{n-1}^{2}L_{n+2(N-k)}^{1}L_{n+3(N-k)}^{1}L_{n+4(N-k)}^{1}\nonumber\\
&&+76L_{n}^{2}L_{n+2(N-k)}^{1}L_{n+3(N-k)}^{1}L_{n+4(N-k)}^{1}
+63L_{n}^{2}L_{n+1+2(N-k)}^{1}L_{n+3(N-k)}^{1}L_{n+4(N-k)}^{1}\nonumber\\
&&+64L_{n}^{2}L_{n+1+2(N-k)}^{1}L_{n+1+3(N-k)}^{1}L_{n+4(N-k)}^{1}
+25L_{n}^{2}L_{n+1+2(N-k)}^{1}L_{n+1+3(N-k)}^{1}L_{n+1+4(N-k)}^{1}\nonumber\\
&&+125L_{n}^{1}L_{n+N-k}^{1}L_{n+2(N-k)}^{1}
L_{n+3(N-k)}^{1}L_{n+4(N-k)}^{1}
\label{quin}    
\end{eqnarray}
 We can see combinatorial characteristics 
of coefficients described in (\ref{poly}) also in this case 
and we assume these coefficients are true. 
Then we determined the remaining unknown coefficients of true recursive formula 
using the following method.

 We first obtain some linear relations 
between them using Conjecture 2. Then, now that we have obtained 
recursive formulas 
for $d\leq 4$ curves and assume Conjecture 4, successive application of 
recursive formula from $N\geq 2k$ region (in this region, what we need 
is only the Schubert numbers !) to $N=k$ region results in linear function 
of the remaining unknown coefficients. Then from information 
of coefficients of hypergeometric series $a_{d}$ and $b_{d}$ in Conjecture 3, 
we can 
obtain infinite number of linear relations on them, varying $N$. 
The final result we obtained is the following.
\begin{eqnarray}
&&L^{N,k,5}_{n} \nonumber\\
&=&\frac{24}{625}L_{n-4}^{5}+\frac{154}{625}L_{n-3}^{5}
+\frac{269}{625}L_{n-2}^{5}+\frac{154}{625}L_{n-1}^{5}
+\frac{24}{625}L_{n}^{5} \nonumber\\
&&+\frac{6}{125}L_{n-3}^{1}L_{n-3+N-k}^{4}
+\frac{3}{50}L_{n-2}^{1}L_{n-3+N-k}^{4}
+\frac{3}{40}L_{n-1}^{1}L_{n-3+N-k}^{4}\nonumber\\
&&+\frac{3}{32}L_{n}^{1}L_{n-3+N-k}^{4}
+\frac{37}{125}L_{n-2}^{1}L_{n-2+N-k}^{4}
+\frac{71}{200}L_{n-1}^{1}L_{n-2+N-k}^{4}\nonumber\\
&&+\frac{17}{40}L_{n}^{1}L_{n-2+N-k}^{4}
+\frac{58}{125}L_{n-1}^{1}L_{n-1+N-k}^{4}
+\frac{393}{800}L_{n}^{1}L_{n-1+N-k}^{4}\nonumber\\
&&+\frac{24}{125}L_{n}^{1}L_{n+N-k}^{4}\nonumber\\
&&+\frac{8}{125}L_{n-3}^{2}L_{n-2+2(N-k)}^{3}+
\frac{8}{75}L_{n-2}^{2}L_{n-2+2(N-k)}^{3}
+\frac{8}{45}L_{n-1}^{2}L_{n-2+2(N-k)}^{3}\nonumber\\
&&+\frac{1}{9}L_{n}^{2}L_{n-2+2(N-k)}^{3} 
+\frac{46}{125}L_{n-2}^{2}L_{n-1+2(N-k)}^{3}
+\frac{122}{225}L_{n-1}^{2}L_{n-1+2(N-k)}^{3}\nonumber\\
&&+\frac{29}{90}L_{n}^{2}L_{n-1+2(N-k)}^{3}+
\frac{59}{125}L_{n-1}^{2}L_{n+2(N-k)}^{3}
+\frac{6}{25}L_{n}^{2}L_{n+2(N-k)}^{3}\nonumber\\
&&+\frac{12}{125}L_{n}^{2}L_{n+1+2(N-k)}^{3}\nonumber\\
&&+\frac{12}{125}L_{n-3}^{3}L_{n-1+3(N-k)}^{2}+
\frac{6}{25}L_{n-2}^{3}L_{n-1+3(N-k)}^{2}
+\frac{29}{90}L_{n-1}^{3}L_{n-1+3(N-k)}^{2}\nonumber\\
&&+\frac{1}{9}L_{n}^{3}L_{n-1+3(N-k)}^{2}
+\frac{59}{125}L_{n-2}^{3}L_{n+3(N-k)}^{2}
+\frac{122}{225}L_{n-1}^{3}L_{n+3(N-k)}^{2}\nonumber\\
&&+\frac{8}{45}L_{n}^{3}L_{n+3(N-k)}^{2}
+\frac{46}{125}L_{n-1}^{3}L_{n+1+3(N-k)}^{2}
+\frac{8}{75}L_{n}^{3}L_{n+1+3(N-k)}^{2}\nonumber\\
&&+\frac{8}{125}L_{n}^{3}L_{n+2+3(N-k)}^{2}\nonumber\\
&&+\frac{24}{125}L_{n-3}^{4}L_{n+4(N-k)}^{1}
+\frac{393}{800}L_{n-2}^{4}L_{n+4(N-k)}^{1}
+\frac{17}{40}L_{n-1}^{4}L_{n+4(N-k)}^{1}\nonumber\\
&&+\frac{3}{32}L_{n}^{4}L_{n+4(N-k)}^{1} 
+\frac{58}{125}L_{n-2}^{4}L_{n+1+4(N-k)}^{1}
+\frac{71}{200}L_{n-1}^{4}L_{n+1+4(N-k)}^{1}\nonumber\\
&&+\frac{3}{40}L_{n}^{4}L_{n+1+4(N-k)}^{1}
+\frac{37}{125}L_{n-1}^{4}L_{n+2+4(N-k)}^{1}
+\frac{3}{50}L_{n}^{4}L_{n+2+4(N-k)}^{1}\nonumber\\
&&+\frac{6}{125}L_{n}^{4}L_{n+3+4(N-k)}^{1}\nonumber\\
&&+\frac{2}{25}L_{n-2}^{1}L_{n-2+N-k}^{1}L_{n-2+2(N-k)}^{3}
+\frac{1}{10}L_{n-1}^{1}L_{n-2+N-k}^{1}L_{n-2+2(N-k)}^{3}\nonumber\\
&&+\frac{1}{8}L_{n}^{1}L_{n-2+N-k}^{1}L_{n-2+2(N-k)}^{3}
+\frac{2}{15}L_{n-1}^{1}L_{n-1+N-k}^{1}L_{n-2+2(N-k)}^{3}\nonumber\\
&&+\frac{1}{6}L_{n}^{1}L_{n-1+N-k}^{1}L_{n-2+2(N-k)}^{3}
+\frac{2}{9}L_{n}^{1}L_{n+N-k}^{1}L_{n-2+2(N-k)}^{3}\nonumber\\
&&+\frac{11}{25}L_{n-1}^{1}L_{n-1+N-k}^{1}L_{n-1+2(N-k)}^{3}
+\frac{21}{40}L_{n}^{1}L_{n-1+N-k}^{1}L_{n-1+2(N-k)}^{3}\nonumber\\
&&+\frac{29}{45}L_{n}^{1}L_{n+N-k}^{1}L_{n-1+2(N-k)}^{3}
+\frac{12}{25}L_{n}^{1}L_{n+N-k}^{1}L_{n+2(N-k)}^{3}\nonumber\\
&&+\frac{6}{25}L_{n-2}^{1}L_{n-2+N-k}^{3}L_{n+4(N-k)}^{1}
+\frac{3}{10}L_{n-1}^{1}L_{n-2+N-k}^{3}L_{n+4(N-k)}^{1}\nonumber\\
&&+\frac{3}{8}L_{n}^{1}L_{n-2+N-k}^{3}L_{n+4(N-k)}^{1}
+\frac{23}{40}L_{n-1}^{1}L_{n-1+N-k}^{3}L_{n+4(N-k)}^{1}\nonumber\\
&&+\frac{2}{3}L_{n}^{1}L_{n-1+N-k}^{3}L_{n+4(N-k)}^{1}
+\frac{3}{8}L_{n}^{1}L_{n+N-k}^{3}L_{n+4(N-k)}^{1}\nonumber\\
&&+\frac{13}{25}L_{n-1}^{1}L_{n-1+N-k}^{3}L_{n+1+4(N-k)}^{1}
+\frac{23}{40}L_{n}^{1}L_{n-1+N-k}^{3}L_{n+1+4(N-k)}^{1}\nonumber\\
&&+\frac{3}{10}L_{n}^{1}L_{n+N-k}^{3}L_{n+1+4(N-k)}^{1}
+\frac{6}{25}L_{n}^{1}L_{n+N-k}^{3}L_{n+2+4(N-k)}^{1}\nonumber\\
&&+\frac{12}{25}L_{n-2}^{3}L_{n+3(N-k)}^{1}L_{n+4(N-k)}^{1}
+\frac{29}{45}L_{n-1}^{3}L_{n+3(N-k)}^{1}L_{n+4(N-k)}^{1}\nonumber\\
&&+\frac{2}{9}L_{n}^{3}L_{n+3(N-k)}^{1}L_{n+4(N-k)}^{1}
+\frac{21}{40}L_{n-1}^{3}L_{n+1+3(N-k)}^{1}L_{n+4(N-k)}^{1}\nonumber\\
&&+\frac{1}{6}L_{n}^{3}L_{n+1+3(N-k)}^{1}L_{n+4(N-k)}^{1}
+\frac{1}{8}L_{n}^{3}L_{n+2+3(N-k)}^{1}L_{n+4(N-k)}^{1}\nonumber\\
&&+\frac{11}{25}L_{n-1}^{3}L_{n+1+3(N-k)}^{1}L_{n+1+4(N-k)}^{1}
+\frac{2}{15}L_{n}^{3}L_{n+1+3(N-k)}^{1}L_{n+1+4(N-k)}^{1}\nonumber\\
&&+\frac{1}{10}L_{n}^{3}L_{n+2+3(N-k)}^{1}L_{n+1+4(N-k)}^{1}
+\frac{2}{25}L_{n}^{3}L_{n+2+3(N-k)}^{1}L_{n+2+4(N-k)}^{1}\nonumber\\
&&+\frac{3}{25}L_{n-2}^{1}L_{n-2+N-k}^{2}L_{n-1+3(N-k)}^{2}
+\frac{3}{20}L_{n-1}^{1}L_{n-2+N-k}^{2}L_{n-1+3(N-k)}^{2}\nonumber\\
&&+\frac{3}{16}L_{n}^{1}L_{n-2+N-k}^{2}L_{n-1+3(N-k)}^{2}
+\frac{3}{10}L_{n-1}^{1}L_{n-1+N-k}^{2}L_{n-1+3(N-k)}^{2}\nonumber\\
&&+\frac{3}{8}L_{n}^{1}L_{n-1+N-k}^{2}L_{n-1+3(N-k)}^{2}
+\frac{1}{3}L_{n}^{1}L_{n+N-k}^{2}L_{n-1+3(N-k)}^{2}\nonumber\\
&&+\frac{14}{25}L_{n-1}^{1}L_{n-1+N-k}^{2}L_{n+3(N-k)}^{2}
+\frac{53}{80}L_{n}^{1}L_{n-1+N-k}^{2}L_{n+3(N-k)}^{2}\nonumber\\
&&+\frac{8}{15}L_{n}^{1}L_{n+N-k}^{2}L_{n+3(N-k)}^{2}
+\frac{8}{25}L_{n}^{1}L_{n+N-k}^{2}L_{n+1+3(N-k)}^{2}\nonumber\\
&&+\frac{4}{25}L_{n-2}^{2}L_{n-1+2(N-k)}^{1}L_{n-1+3(N-k)}^{2}
+\frac{4}{15}L_{n-1}^{2}L_{n-1+2(N-k)}^{1}L_{n-1+3(N-k)}^{2}\nonumber\\
&&+\frac{1}{6}L_{n}^{2}L_{n-1+2(N-k)}^{1}L_{n-1+3(N-k)}^{2}
+\frac{2}{5}L_{n-1}^{2}L_{n+2(N-k)}^{1}L_{n-1+3(N-k)}^{2}\nonumber\\
&&+\frac{1}{4}L_{n}^{2}L_{n+2(N-k)}^{1}L_{n-1+3(N-k)}^{2}
+\frac{1}{6}L_{n}^{2}L_{n+1+2(N-k)}^{1}L_{n-1+3(N-k)}^{2}\nonumber\\
&&+\frac{17}{25}L_{n-1}^{2}L_{n+2(N-k)}^{1}L_{n+3(N-k)}^{2}
+\frac{2}{5}L_{n}^{2}L_{n+2(N-k)}^{1}L_{n+3(N-k)}^{2}\nonumber\\
&&+\frac{4}{15}L_{n}^{2}L_{n+1+2(N-k)}^{1}L_{n+3(N-k)}^{2}
+\frac{4}{25}L_{n}^{2}L_{n+1+2(N-k)}^{1}L_{n+1+3(N-k)}^{2}\nonumber\\
&&+\frac{8}{25}L_{n-2}^{2}L_{n-1+2(N-k)}^{2}L_{n+4(N-k)}^{1}
+\frac{8}{15}L_{n-1}^{2}L_{n-1+2(N-k)}^{2}L_{n+4(N-k)}^{1}\nonumber\\
&&+\frac{1}{3}L_{n}^{2}L_{n-1+2(N-k)}^{2}L_{n+4(N-k)}^{1}
+\frac{53}{80}L_{n-1}^{2}L_{n+2(N-k)}^{2}L_{n+4(N-k)}^{1}\nonumber\\
&&+\frac{3}{8}L_{n}^{2}L_{n+2(N-k)}^{2}L_{n+4(N-k)}^{1}
+\frac{3}{16}L_{n}^{2}L_{n+1+2(N-k)}^{2}L_{n+4(N-k)}^{1}\nonumber\\
&&+\frac{14}{25}L_{n-1}^{2}L_{n+2(N-k)}^{2}L_{n+1+4(N-k)}^{1}
+\frac{3}{10}L_{n}^{2}L_{n+2(N-k)}^{2}L_{n+1+4(N-k)}^{1}\nonumber\\
&&+\frac{3}{20}L_{n}^{2}L_{n+1+2(N-k)}^{2}L_{n+1+4(N-k)}^{1}
+\frac{3}{25}L_{n}^{2}L_{n+1+2(N-k)}^{2}L_{n+2+4(N-k)}^{1}\nonumber\\
&&+\frac{1}{5}L_{n-1}^{1}L_{n-1+N-k}^{1}L_{n-1+2(N-k)}^{1}L_{n-1+3(N-k)}^{2}
+\frac{1}{4}L_{n}^{1}L_{n-1+N-k}^{1}L_{n-1+2(N-k)}^{1}
L_{n-1+3(N-k)}^{2}\nonumber\\
&&+\frac{1}{3}L_{n}^{1}L_{n+N-k}^{1}L_{n-1+2(N-k)}^{1}L_{n-1+3(N-k)}^{2}
+\frac{1}{2} L_{n}^{1}L_{n+N-k}^{1}L_{n+2(N-k)}^{1}
L_{n-1+3(N-k)}^{2}\nonumber\\
&&+\frac{4}{5}L_{n}^{1}L_{n+N-k}^{1}L_{n+2(N-k)}^{1}L_{n+3(N-k)}^{2}
+\frac{2}{5}  L_{n-1}^{1}L_{n-1+N-k}^{1}L_{n-1+2(N-k)}^{2}
L_{n+4(N-k)}^{1}\nonumber\\
&&+\frac{1}{2}L_{n}^{1}L_{n-1+N-k}^{1}L_{n-1+2(N-k)}^{2}L_{n+4(N-k)}^{1}
+\frac{2}{3} L_{n}^{1}L_{n+N-k}^{1}L_{n-1+2(N-k)}^{2}
L_{n+4(N-k)}^{1}\nonumber\\
&&+\frac{3}{4} L_{n}^{1}L_{n+N-k}^{1}L_{n+2(N-k)}^{2}L_{n+4(N-k)}^{1}
+\frac{3}{5}  L_{n}^{1}L_{n+N-k}^{1}L_{n+2(N-k)}^{2}
L_{n+1+4(N-k)}^{1}\nonumber\\
&&+\frac{3}{5}L_{n-1}^{1}L_{n-1+N-k}^{2}L_{n+3(N-k)}^{1}L_{n+4(N-k)}^{1}
+\frac{3}{4}L_{n}^{1}L_{n-1+N-k}^{2}L_{n+3(N-k)}^{1}L_{n+4(N-k)}^{1}\nonumber\\
&&+\frac{2}{3}L_{n}^{1}L_{n+N-k}^{2}L_{n+3(N-k)}^{1}L_{n+4(N-k)}^{1}
+\frac{1}{2}L_{n}^{1}L_{n+N-k}^{2}L_{n+1+3(N-k)}^{1}L_{n+4(N-k)}^{1}\nonumber\\
&&+\frac{2}{5}L_{n}^{1}L_{n+N-k}^{2}L_{n+1+3(N-k)}^{1}L_{n+1+4(N-k)}^{1}
+\frac{4}{5}L_{n-1}^{2}L_{n+2(N-k)}^{1}L_{n+3(N-k)}^{1}
L_{n+4(N-k)}^{1}\nonumber\\
&&+\frac{1}{2}L_{n}^{2}L_{n+2(N-k)}^{1}L_{n+3(N-k)}^{1}L_{n+4(N-k)}^{1}
+\frac{1}{3}L_{n}^{2}L_{n+1+2(N-k)}^{1}L_{n+3(N-k)}^{1}
L_{n+4(N-k)}^{1}\nonumber\\
&&+\frac{1}{4}L_{n}^{2}L_{n+1+2(N-k)}^{1}L_{n+1+3(N-k)}^{1}L_{n+4(N-k)}^{1}
+\frac{1}{5}L_{n}^{2}L_{n+1+2(N-k)}^{1}L_{n+1+3(N-k)}^{1}
L_{n+1+4(N-k)}^{1}\nonumber\\
&&+L_{n}^{1}L_{n+N-k}^{1}L_{n+2(N-k)}^{1}
L_{n+3(N-k)}^{1}L_{n+4(N-k)}^{1}
\label{quin2}    
\end{eqnarray}
This formula correctly predicts $L_{m}^{N,k,5}$ in $N-k\geq 1$ 
region and reproduce coefficients of hypergeometric functions in 
$N-k=0$ region if we start from $N\geq 2k$ region and input 
Schubert numbers.
Of course, we can inductively obtain recursive formula for curves 
of higher degree using the same method, but general structure 
of the coefficients that appear in recursive formula is still 
an open problem. 
\section{Conclusion}
In this paper, we proved the fact that correlation 
functions of hypersurfaces $M_{N}^{k}$ in $CP^{N-1}$ ($N\geq k$) 
can be written as polynomials of finite number of integers $L_{m}^{k}$
up to degree 3.
In quintic case, these numbers are 1345, 770, 120. We cannot 
tell how these results are used in the future, but in 
proving this, we found the recursion relations that is invariant 
in $c_{1}(M_{N}^{k})\geq 2$ case produces ``bare'' B-model 
or ``bare'' coordinates of deformation of complex structure
of mirror manifold of $M_{k}^{k}$. This completely agrees with the 
results of Givental, which  saids that in $c_{1}(M_{N}^{k})\geq 2$ case, 
sigma models on $(M_{N}^{k})$ can be solved with hypergeometric 
series without coordinate transformation i.e., (bare deformation 
parameter is good coordinate of A-model) and that in Calabi-Yau case, 
we have to translate the bare coordinate by mirror map. 
In sum, we can say B-model as toric quantum cohomology 
compatible with toric compactification of moduli space of pure matter
theory. And in Calabi-Yau case, we have to introduce mirror 
map to compensate for the gap between toric compactification of 
moduli space 
of pure matter theory and exact moduli space. 
These conclusion agrees with the argument of \cite{mp}. 
Maybe application of complete intersections 
in $CP^{N-1}$ can be achieved by changing the input integers $L_{m}^{k}$.
We also have to search for the generalization 
of specialization arguments to the case of weighted projective space. 
In this case, we would find toric structure of quantum cohomology 
ring by construction of recursion relations.   

Our last step in discussion of Calabi-Yau hypersurfaces in $CP^{N-1}$
is construction of correspondence between correction terms and 
boundary parts of toric compactifications of moduli space. 
 
{\bf Acknowledgment}
A.C. thanks to A.B.~Givental, J.~Lewis, C.~Peters, S.A.Str$\o$mme,
 G.~Tian, Tyurin. 
 
M.J. thanks to T.~Eguchi, K.~Hori, A.~Matsuo, M.~Kobayashi, 
members of Algebraic Geometry Groups in Dept. of Mathematical Science
in Tokyo Univ, M.~Nagura and Y.~Sun for discussions and kind encouragement. 
A.C. has been partially supported by 
Science Project Geom. of Alg. Var.
n. 0-198-SC1 , and by fundings from 
M.U.R.S.T. and G.N.S.A.G.A. (C.N.R.)
Italy.
M.J has been supported by grant of Japanese Ministry of Education. 


\smallskip
\begin{thebibliography}{99}
\bibitem{beauville}A.Beauville
\newblock{\em Quantum cohomology of complete intersections,}
\newblock Mathematical, Physics Analysis and Geometry  168 (1995), 384-398.
\bibitem{bert}A.Bertram.
\newblock{\em Quantum Schubert Calculus}
\newblock To appear in Advances in Mathematics
\bibitem{blmur}Bloch-Murre
\newblock{\em On the Chow group of certain types of Fano
threefolds,}
\newblock{Compositio Math. {\bf 39}, 47-105, 1979}
\bibitem{candelas1} P.~Candelas and X.~de la Ossa,
Nucl.~Phys. {\bf B355} (1991) 415.
\bibitem{candelas2} P.~Candelas, X.~de la Ossa, P.~Green and L.~Parkes,
Phys.~Lett. {\bf 258B} (1991) 118; Nucl.~Phys. {\bf B359} (1991) 21.
\bibitem{collino}A.Collino.
\newblock{\em Some computations on the quantum cohomology
algebra of a Fano hypersurface,}
\newblock Informal draft (1996).
\bibitem{dub}B.Dubrovin.
\newblock{\em The geometry of 2D topological field theories,}
\newblock in Integral systems and quantum groups, (LNM
1620, Springer-Verlag 1996) 120-348.
\bibitem{ES} Ellingsrud, S.A.Str$\o$mme
\newblock{\em Bott's formula and enumerative geometry,}
\newblock Jour. AMS 9 (1996),n.1, 175-193.
\bibitem{Fu1} W.~Fulton
\newblock{\em Intersection Theory}
\newblock{Ergebnisse der Math. und ihrer Grenzgebiete 3. Folge Band 2
Springer-Verlag, 1984}
\bibitem{Fu2} Fulton and  Pandharipande.
\newblock{\em Notes on stable maps
and quantum cohomology,}
\newblock in Proceedings of symposia in pure mathematics: Algebraic
geometry Santa Cruz 1995, (J. Kollar, R. Lazarsfeld, D. Morrison eds.) Volume 62,
Part 2,45-96. (American Mathematical Society).
\bibitem{givental}A.B.Givental
\newblock{\em Equivariant Gromov-Witten Invariants,}
\newblock Internat. Math. Res.Notices 13 (1996),613--663.
\bibitem{h.d.m.}B.R. Greene, D.R. Morrison and
M.R. Plesser
\newblock{\em Mirror Manifolds in Higher Dimension,}
\newblock{Commun. Math. Phys. 173 (1995) 559-598}
\bibitem{griffiths}P.~Griffiths and J.~Harris, {\em Principles of Algebraic
Geometry}, ( Wiley, 1978 )
\bibitem{ho}S.Hosono, A.Klemn, S.Theisen and S.T.Yau
\newblock{\em Mirror Symmetry, Mirror Map and Applications
to Calabi-Yau Hypersurfaces,}
\newblock{Commun.Math.Phys. 167 (1995) 301-350}
\bibitem{fusion}K.Intriligator.
\newblock{\em Fusion residues,}
\newblock{Modern Physics letters A6 (1991), Number 38,
pp. 3543-3556.}
\bibitem{jin}M.Jinzenji
\newblock{\em On Quantum Cohomology Rings for Hypersurfaces in $CP^{N-1}$,}
\newblock J.Math.Phys. 38 (1997) 5775-5802.
\bibitem{j}M.Jinzenji.
\newblock{\em Construction of Free Energy of Calabi-Yau Manifold
embedded in $CP^{N-1}$ via Torus Actions,}
\newblock Int.J.Mod.Phys. A12 (1997) 5775-5802
\bibitem{mnmj}M.Jinzenji and M.Nagura
\newblock{\em Mirror Symmetry and An Exact Calculation of
$N-2$ point Correlation Function on Calabi-Yau Manifold embedded
in $CP^{N-1}$,}
\newblock Int.J.Mod.Phys. A11 (1996) 171-202
\bibitem{ke}S.Keel.
\newblock{\em Intersection theory of moduli spaces of n-stable
pointed curves of genus zero,}
\newblock Trans. Ams, 330 (1992), 545-574.
\bibitem{tor} M.Kontsevich.
\newblock{\em Enumeration of Rational Curves via Torus Actions,}
\newblock In: The moduli space of curves, R.Dijkgraaf,
C.Faber, G.van der Geer (Eds.), Progress in Math., v.129,
Birkh\"auser, 1995, 335-368.
\bibitem{km} M.Kontsevich , Y.Manin.
\newblock{\em Gromov--Witten Classes, Quantum Cohomology,
and Enumerative
 Geometry,}
\newblock Commun.Math.Phys.164 (1994) 525-562
\bibitem{L} J.~Lewis
\newblock{\em The cylinder correspondence for hypersurfaces
of degree $n$ in $P^{n}$.}
\newblock{American Journal of Mathematics, {\bf 110}, 77-114}
\bibitem{ruan}J.Li, G.Tian.
\newblock{\em Quantum Cohomology of Homogeneous Varieties,}
\newblock alg/geom/9504009
\bibitem{mp}D.R.Morrison and M.R.Plesser
\newblock{\em Summing the Instantons: Quantum Cohomology and
Mirror Symmetry in Toric Varieties,}
\newblock Nucl.Phys. B440 (1995) 279-354
\bibitem{nag} M.Nagura and K.Sugiyama,
\newblock{\it Mirror Symmetry of K3 Surface,}
\newblock Int.J.Mod.Phys. A10 (1995) 233
\bibitem{PP} U. Persson, C.Peters.
\newblock{\em Some aspects of the topology of algebraic surfaces,}
Israel Mathematical conference Proceedings Vol 9, (1996), 377-392
\bibitem{ruantian} Y.~Ruan, G.~Tian
\newblock{\em A mathematical theory of quantum cohomology,}
\newblock{J.Diff.Geom.42 \quad no.2\quad 1995}
\bibitem{tian}G.~Tian
\newblock{\em Quantum cohomology and its associativity,}
\newblock{Proc. of 1st Current Developments in Math.,
Cambridge 1995.}
\bibitem{Ty} Tjurin, A. N.
\newblock{\em Five lectures on three-dimensional varieties, (Russian)}
\newblock{Uspehi Mat. Nauk 27 (1972), no. 5, (167), 3--50.}
\bibitem{va}C.Vafa.
\newblock{\em Topological Mirrors and Quantum Rings,}
\newblock hep-th/9111017
\bibitem{witten1}E.Witten
\newblock{\em Mirror Manifolds and Topological Field Theory,}
\newblock in Essays on Mirror Manifolds, ed. S.-T.Yau (Int. Press.
Co.,Hong Kong, 1992) 120-180
\end{thebibliography}
\end{document}